\documentclass[useAMS,usenatbib]{mn2e}

\usepackage{graphicx}
\usepackage{aas_macros}
\usepackage{amsfonts,amssymb}
\usepackage[
    a4paper,
    plainpages=false,
    breaklinks=true,
    bookmarks=true,
    bookmarksopen=true,
    bookmarksopenlevel=1
]{hyperref}

\def\lesssim{\mathrel{\hbox{\rlap{\hbox{\lower4pt\hbox{$\sim$}}}\hbox{$<$}}}}
\def\gtrsim{\mathrel{\hbox{\rlap{\hbox{\lower4pt\hbox{$\sim$}}}\hbox{$>$}}}}

\title[Redshift Evolution of Cold Gas Mass Function]
      {The Redshift Evolution of the Mass Function of Cold Gas in Hierarchical
        Galaxy Formation Models}
      
      \author[C.~Power, C.~M.~Baugh \& C.~G.~Lacey]
	     { C.~Power\thanks{chris.power@astro.le.ac.uk}$^{1,2}$,
               C.~M.~Baugh$^3$ \& C.~G.~Lacey$^3$\\
               $^1$ Department of Physics \& Astronomy, University of 
               Leicester, Leicester LE1 7RH, United Kingdom\\
	       $^2$ Centre for Astrophysics and Supercomputing,
	       Swinburne University of Technology,
	       PO Box 218, Hawthorn, 3122, Victoria, Australia\\
               $^3$ Institute for Computational Cosmology, University 
	       of Durham, South Road, Durham DH1 3LE, United Kingdom}
	     
\begin{document}

\date{}

\pagerange{\pageref{firstpage}--\pageref{lastpage}} \pubyear{2010}

\maketitle

\label{firstpage}

\begin{abstract}

Accurately predicting how the cosmic abundance of neutral hydrogen
evolves with redshift is a challenging problem facing modellers of galaxy 
formation. We investigate the predictions of four currently favoured
semi-analytical galaxy formation models applied to the Millennium 
simulation for the mass function of cold neutral gas (atomic and molecular)
in galaxies as a function of redshift, and we use these predictions to 
construct number counts for the next generation of all-sky neutral atomic 
hydrogen (HI) surveys. Despite the different 
implementations of the physical ingredients of galaxy formation, we find 
that the model predictions are broadly consistent with one another; the key
differences reflect how the models treat AGN feedback and how the 
timescale for star formation evolves with redshift. The models produce 
mass functions of cold gas in galaxies that are generally in good agreement
 with HI surveys at $z$=0. Interestingly we find that these mass functions do 
not evolve significantly with redshift. Adopting a simple conversion factor
for cold gas mass to HI mass that we apply to all galaxies at all redshifts, 
we derive mass functions of HI in galaxies from the predicted mass functions 
of cold gas, which we use to predict the number counts of sources likely to 
be detected by HI surveys on next generation radio telescopes such as the 
Square Kilometre Array and its pathfinders. We find the number counts peak 
at $\sim 4\times 10^3/4\times 10^4/3\times 10^5$ galaxies per square degree at 
$z\!\sim$ 0.1/0.2/0.5 for a year long HI hemispheric survey on a 
1\%/10\%/100\% SKA with a 30 square degree field of view, corresponding to an 
integration time of 12 hours. On a full SKA with a 200 square degree field of 
view (equivalent to an integration time of 80 hours) the number counts peak 
at $5\times 10^5$ galaxies per square degree at $z\!\sim$ 0.6. We 
show also how adopting a conversion factor for cold gas mass to HI mass that 
varies from galaxy to galaxy impacts on number counts. In addition, we 
examine how the typical angular sizes of galaxies vary with redshift. These 
decline strongly with increasing redshift at $z\!\lesssim\!0.5$ and more gently
at $z\!\gtrsim\!0.5$; the median angular size varies between $5''$ and $10''$ 
at $z$=0.1, $0.5''$ and $3''$ at $z$=1 and $0.2''$ and $1''$ at $z$=3 for
galaxies with HI masses in excess of $10^9 h^{-1} \rm M_{\odot}$, depending on 
the precise model. Taken together, these results make clear that 
forthcoming HI surveys will provide important and powerful tests of
theoretical galaxy formation models.

\end{abstract}


\section{Introduction}
\label{sec:intro}

Neutral gas, predominantly atomic hydrogen (HI) along with molecular 
hydrogen (H$_2$) and helium (He), plays a fundamental role in galaxy formation,
 principally as the raw material from which stars are made. At any given 
time the fraction of a galaxy's mass that is in the form of HI
will be determined by the competing rates at which it is depleted (by, 
for example, star formation, photo-ionisation and expulsion via winds) 
and replenished (by, for example, recombination and accretion from the 
galaxy's surroundings). These processes are integral to any theory of 
galaxy formation and so we expect that understanding how the HI properties 
of galaxies vary with redshift and environment will provide us with important 
insights into how galaxies form.

Our knowledge of HI in galaxies derives from radio observations of the 
rest-frame 21-cm emission line, which allows us to measure the density, 
temperature and velocity distribution of HI along our line of sight. Thanks to 
surveys such as HIPASS \citep[HI Parkes All-Sky Survey; 
see][]{meyer.etal.2004} and more recently ALFALFA \citep[Arecibo Legacy 
Fast ALFA Survey; see][]{giovanelli.etal.2005}, we have a good understanding 
of the HI properties of galaxies in the low-redshift Universe, at
$z \lesssim 0.05$. For example, HIPASS has revealed that most HI is 
associated with galaxies and that the galaxy population detected in 
21-cm emission is essentially the same as that seen at optical and infrared 
wavelengths but weighted towards gas-rich systems, which tend to be late-type 
\citep{meyer.etal.2004}. Furthermore, HIPASS data have allowed accurate
measurement of the local HI mass function of galaxies (the number density of 
galaxies with a given HI mass per unit comoving volume) and $\Omega_{\rm HI}$ 
(the global HI mass density in units of the present-day critical density 
$\rho_{\rm crit}=3H_0^2/8\pi\,G$). \citet{zwaan.etal.2005} found that the
local HI mass function is well described by a Schechter function and they 
estimated the local cosmic HI mass density to be $\Omega_{\rm HI}=3.5 \pm 0.4 
\pm 0.4 \times 10^{-4}\,h_{75}^{-1}$ (with random and systematic uncertainties 
at the $68\%$ confidence limit), assuming a dimensionless Hubble parameter 
of $h_{75}$=$h/0.75$=1. This is approximately $1/10^{\rm th}$ the 
value of cosmic stellar mass density $\Omega_{\ast}$ at $z$=0 
\citep[cf.][]{cole.etal.2001}

In contrast, we know comparatively little about HI in galaxies at higher 
redshifts (i.e. $z \gtrsim 0.05$). This is because detecting the rest frame 
21-cm emission from individual galaxies has required too great a sensitivity 
for reasonable observing times\footnote{Although in a handful of cases it has 
been possible to use a stacking technique to measure rest frame 21-cm emission 
by co-adding the signal from multiple galaxies using their observed optical 
positions and redshifts; see \cite{zwaan.2000}, \cite{2001A&A...372..768C} and 
\cite{lah.etal.2007,lah.etal.2009}.}. There are estimates of $\Omega_{\rm HI}$ 
at high redshifts ($z \gtrsim 1.5$) but these have been deduced from QSO 
absorption-line systems and imply that $\Omega_{\rm HI} \simeq 10^{-3}$ 
\citep[e.g.][]{peroux.etal.2003,prochaska.etal.2005,rao.etal.2006}. 
However, this situation will change dramatically over the next decade with the 
emergence of a series of next 
generation radio telescopes, culminating in the Square Kilometre Array
(SKA) that is expected to see first light by about 2020. The SKA will have 
sufficient sensitivity and angular resolution to map HI in galaxies out to 
redshifts $z \gtrsim 3$ \cite[see, for example][]{blake.etal.2007, braun.2007}.
On a shorter timescale a variety of SKA pathfinders such as ASKAP 
\citep[the Australian SKA Pathfinder; see][]{askap.science.2008}, MeerKAT 
\citep[the Karoo Array Telescope; see][]{meerkat.2007} and APERTIF 
\citep[APERture Tile In Focus; see][]{apertif.2008} will carry out HI surveys 
that, in some cases, will probe the properties of galaxies out to $z \sim 1$.\\

Because we have so little data on HI in galaxies beyond the local Universe,
the results of HI surveys on next generation radio telescopes
will have a profound impact on our understanding of galaxy formation and 
evolution. For example, we have compelling evidence that the cosmic star 
formation rate density has decreased by an order of magnitude since $z\sim 1$ 
\citep{madau.etal.1996,hopkins.2004}, yet we know little about how the HI mass
in galaxies evolved over the same period. This is precisely the kind of 
question we can hope to answer with forthcoming HI surveys. 
For this reason it is both timely and important to take stock of 
what theoretical galaxy formation models tell us about how quantities such 
as $\Omega_{\rm HI}$ and the HI mass function vary with redshift and 
environment. Not only can such predictions help us to interpret the physical 
significance of observational data, they can also provide important 
input into the design of new radio telescopes.

The primary aim of this paper is to explore the predictions of currently 
favoured \emph{semi-analytical} galaxy formation models 
\citep[cf.][]{cole.etal.2000,baugh.2006}
for the properties of cold gas in galaxies as a function of redshift in a 
$\Lambda$ Cold Dark Matter ($\Lambda$CDM) universe. In particular we 
investigate the redshift variation of the cold gas mass function of galaxies 
and $\Omega_{\rm cold}$ (i.e. the cold gas mass density parameter) 
in the Millennium simulation \citep{springel.etal.2005}. In addition we 
convert the predicted cold gas masses to HI masses and we use the resulting 
HI mass functions, along with predictions for the radii
and rotation speeds of galactic discs, to predict the number counts of sources 
one might expect to recover from HI surveys on a radio telescope with a
collecting area of $1\%$/$10\%$/$100\%$ of the full SKA. 

The secondary aim of this paper is to compare and contrast the predictions 
of four distinct galaxy formation models from the Durham and Munich groups. 
These models incorporate different treatments of the same physical processes 
and in some cases invoke distinct physical processes, for example, AGN heating 
versus supernova-driven super-winds. Therefore it is instructive to assess the 
robustness of the basic predictions and to examine whether or not these 
predictions are consistent between models. This addresses the criticism that 
semi-analytical galaxy formation models lack transparency and the uncertainty 
as to which predictions are robust and which are sensitive to modest changes 
in the model parameters. It is in this context that this work complements in a 
very natural way the study of \citet{obreschkow2009b}. \\ 

The layout of the paper is as follows. In \S\ref{sec:galform} we present an 
overview of the four galaxy formation models we use in this study. In 
\S\ref{sec:results} we examine the basic predictions of these models for the 
evolution of the mass function and global mass density of cold gas between 
$0 \lesssim z \lesssim 2$, and we determine the relationship between cold gas
mass and the circular velocities and scale radii of discs. Based on these 
predictions, in \S\ref{sec:observations} we determine what the implications
are for the number counts of HI sources in future HI surveys
on next generation radio telescopes, discussing in some detail how 
we might convert from cold gas mass to HI mass in 
\S\ref{ssec:convert_cg_to_hi}. Finally, in \S\ref{sec:summary} we summarise 
our results and discuss how future HI surveys will provide a powerful test of 
theoretical models of galaxy formation.

\section{Galaxy Formation Models}
\label{sec:galform}

The evolution of the global cold gas density in the Universe and 
the cold gas content of galaxies depends upon the interplay between 
a number of processes: 

\begin{enumerate}
\item the rate at which gas cools radiatively within dark matter haloes;
\item the rate at which cold gas is accreted in galaxy mergers;
\item the rate at which cold gas is consumed in star formation; 
\item the rate at which gas is reheated or expelled from galaxies by 
  sources of feedback (e.g. photo-ionisation, stellar winds, supernovae, 
  AGN heating, etc..).
\end{enumerate}

\noindent Semi-analytical modelling provides us with the means to 
study the balance between these phenomena in the context of a universe 
in which structure in the dark matter grows hierarchically 
\citep[for a recent review, see][]{baugh.2006}. 
The models refer to cold gas as gas that has cooled radiatively 
from a hot phase to below $10^{4}\rm K$ and is available for star formation. 
The cold gas mass is predominately made up of neutral atomic hydrogen (HI), 
along with molecular hydrogen and helium (see discussion in 
\S\ref{ssec:convert_cg_to_hi}).

Here we consider four different semi-analytical galaxy formation models from 
the Durham and Munich groups. Although the models follow the same basic 
philosophy, the implementations of various processes differ substantially 
between the two groups. We also consider Durham models with different 
physical ingredients. To remove one possible source of difference between 
the models, all the models discussed here adopt the background cosmology 
used in the Millennium simulation \citep[$\Omega_{\rm M}$=0.25,
$\Omega_{\Lambda}$=0.75, $\Omega_{\rm b}$ = 0.045, $\sigma_8$=0.9,
$h$=0.73; cf.][]{springel.etal.2005}.

\subsection{The Models}

We now list the different models considered in this paper, give their 
designation and a very brief description of the main features of each. The 
differences between the models are discussed in more detail later on in this 
section. The first three models are ``Durham'' models, which use 
the {\small GALFORM} code, and the fourth model is the current ``Munich'' 
semi-analytical model. The model designations are those used in the 
Millennium Archive\footnote{The Millennium galaxy archive can be found 
at Durham (http://galaxy-catalogue.dur.ac.uk:8080/Millennium)
or Munich (http://www.g-vo.org/Millennium)} and in the subsequent plots.  

\begin{itemize}

\item The \citet{bower.etal.2006} model (hereafter Bower2006a). In this model,
AGN heating suppresses the formation of bright, massive galaxies by stopping 
the cooling flow in their host dark matter haloes, thereby cutting off the 
supply of cold gas for star formation. This regulation of the cooling flow 
results in a sharp break at the bright end of the luminosity function. 
Bower2006a matches the evolution of the stellar mass function 
inferred from observations
\citep[e.g.][]{2004A&A...424...23F,2005ApJ...619L.131D}, 
the number counts and redshift distribution of extremely red objects 
\citep[][]{gonzalez.etal.2009}
and the abundance of luminous red galaxies \citep[][]{almeida.etal.2008}

\item The \citet{font.etal.2008} model (hereafter Font2008a). This model 
extends Bower2006a with a fundamental change to the cooling model. Motivated 
by the simulations of \citet{mccarthy.etal.2008}, which track the fate of the 
hot gas in haloes after their accretion by more massive objects, Font2008a
assumes that the stripping of hot gas from satellite haloes 
is not completely efficient, contrary  
to the traditional recipe used in semi-analytical models. Instead, the 
satellite halo is assumed to retain some fraction of its hot gas, which 
is determined by its orbit within the larger halo. This gas can cool directly 
onto the satellite rather than the central galaxy in the halo. Font2008a 
gives an improved match to the proportions of red and blue galaxies 
seen in SDSS groups \citep[][]{weinmann.etal.2006a,weinmann.etal.2006b}.

\item The \citet{baugh.etal.2005} model (hereafter Baugh2005M). Baugh2005M 
matches the observed counts and redshifts of sub-mm galaxies and the luminosity 
function of Lyman-break galaxies, as well as observations of the local galaxy 
population, such as the sizes of galaxy discs \citep[cf.][]{gonzalez.etal.2008}
and cold gas mass fractions. In this model, merger-triggered star-bursts make a 
similar contribution to the star formation rate per unit volume at high 
redshift to that from galactic discs. Star-bursts are assumed to have a 
top-heavy stellar initial mass function (IMF), which Baugh et al. argued is 
essential for a hierarchical galaxy formation model to match the sub-mm counts, 
whilst at the same time reproducing observations of local galaxies. 
The formation of bright galaxies is regulated by a supernova driven 
``super-winds'', which expel gas from intermediate mass dark matter haloes 
\citep[see ][]{benson.etal.2003.b}. 

In this paper we implement Baugh2005M in the Millennium simulation. The 
cosmology used in the Millennium simulation is somewhat different to that 
adopted in the original Baugh et~al. model (the former has a matter density of 
$\Omega_{\rm M}=0.25$, a dimensionless Hubble parameter of $h=0.73$ and a 
power spectrum normalisation of $\sigma_8=0.9$, whereas the latter used 
$\Omega_{\rm M}=0.3$, $h=0.7$ and $\sigma_8=0.93$). To reproduce the predictions
of Baugh et~al., we retain the baryon fraction $\Omega_{\rm b}/\Omega_{\rm m}$ 
of the original model, setting $\Omega_{\rm b}=0.033$. The other galaxy 
formation parameters have {\it not} been changed. This model is not available 
in the Millennium Archive. 

\item The \citet{delucia.blaizot.2007} model (hereafter DeLucia2006a). 
DeLucia2006a employs AGN feedback in the ``radio-mode'' to restrict the 
formation of bright galaxies at the present day. This model is a development 
of those introduced by \citet{croton.etal.2006} and 
\citet{delucia.etal.2006}. It enjoys many of the same successes as Bower2006a, 
but, if anything, produces too many stars at high redshift 
\citep[cf.][]{kitzbichler.white.2007}
  
\end{itemize}

\subsection{Halo Identification and Merger Trees}

All of the models use halo merger histories extracted from the 
Millennium simulation, derived from identical group catalogues produced by the 
{\small SubFind} code of \citet{springel.etal.2001}. {\small 
SubFind} identifies distinct groups of particles using the 
``friends-of-friends'' (FOF) algorithm \citep[cf.][]{davis.etal.1985} and then 
resolves each FOF group into self-bound overdensities. These self-bound
overdensities correspond to subhaloes; the most massive subhalo within a 
FOF group is identified with the host dark matter halo while the remaining
lower mass subhaloes within the group correspond to its substructures.
There are as many group catalogues as there are output times and they
are linked across multiple output times to produce merger trees.

Although the input group catalogues are identical, the Durham and Munich 
groups construct their merger trees independently using distinct
algorithms. This means that it is possible to identify the same haloes at a
given output time in the Durham and Munich models but the detailed merging 
histories of these haloes may differ. This difference reflects in
part differences in the working definition of a halo. The Munich models 
track the set of particles that correspond to the most massive subhalo 
within the FOF group across output times \citep[cf.][]{croton.etal.2006}, 
whereas the Durham models track the set of particles that correspond -- in 
general -- to the FOF group \citep[cf.][]{helly.etal.2003,harker.etal.2006}. 
In general, because these FOF groups are modified to avoid situations where 
the groups become prematurely or temporarily linked by bridges of low-density
material \citep[cf.][]{harker.etal.2006,cole.etal.2007}.

The difference also reflects how the host subhaloes of satellites are
treated. In the Durham models, an infalling subhalo is considered a 
satellite galaxy of its more massive host halo once it loses in excess of 
25$\%$ of the mass it had at the time of its accretion and it lies within 
twice its host halo's half-mass radius \citep[cf.][]{harker.etal.2006}. 
Importantly, this subhalo is then treated as a satellite at all subsequent 
times, even if its orbit brings it outside of its host's virial radius at some 
later time. In contrast, the Munich models simply require that the infalling 
satellite lies within the virial radius of its host; if the subhalo's orbit 
takes it beyond the virial radius, it is no longer classed as a satellite 
(John Helly, private communication). Also, the Durham models require that the 
masses of subhaloes must increase \emph{monotonically} with time, whereas
the Munich models allow a subhalo's mass to either increase and decrease with 
time. Finally, the Durham models explicitly tag subhaloes that are satellites
and treat them accordingly, whereas the Munich models do not treat subhaloes
that host satellites any differently from subhaloes that do not.

\subsection{Galaxy Formation Physics}

We now highlight some of the areas in which there are either important 
differences in the implementation of the physics between the models or 
in which different processes have been adopted. For full descriptions of 
each model we refer the reader to the original references given above. 
A comparison of Bower2006a and Baugh2005M is given in
\citet{almeida.etal.2007}; the differences between Bower2006a and 
Font2008a are set out in \citet{font.etal.2008}.

{\it (1) Gas cooling: gas density and cooling radius}.  The models all
assume that gas cools primarily (for the haloes which typically host
galaxies) by two-body collisional processes involving neutral or
ionised atoms. The cooling rate depends upon the composition
(metallicity) and the density of the gas.  Gas is assumed to have
cooled within some cooling radius, which is defined in different ways
in the models. A further timescale that regulates the addition of
cold gas into a galactic disc is the free-fall time. 

The models make different assumptions about the density profile of the 
gas and the cooling radius: 

\vspace{0.25cm}

\noindent (i) DeLucia2006a assumes that the gas follows 
a singular isothermal profile \citep[see ][]{croton.etal.2006}, and the cooling 
radius is defined as the radius at which the cooling time is equal to the
dynamical time of the halo. 

\vspace{0.1cm}

\noindent (ii) Both Bower2006a and Font2008a assume that the hot gas 
density profile follows an isothermal profile with a constant density
core, whose core radius is fixed and scales with the virial radius of
the halo. The cooling radius propagates outwards as a function of
time, reaching a maximum at the radius where the cooling time is equal
to the lifetime of the dark matter halo \citep[see ][]{cole.etal.2000}. In
Font2008a, the cold gas yield is a factor of two higher than that adopted in
Bower2006a, which gives a better match to observed galaxy colours
\citep{gonzalez.etal.2008}.  

\vspace{0.1cm}

\noindent (iii) Baugh2005M assumes that the hot gas follows an isothermal 
profile with a constant density core, whose core radius evolves with time as 
low entropy gas cools \citep[see ][]{cole.etal.2000}. The cooling radius is
defined in the same way as in Bower2006a and Font2008a.

\vspace{0.25cm}

{\it (2) Gas cooling: AGN heating of the hot halo.} 
Bower2006a, Font2008a and DeLucia2006a all modify the cooling flow in massive 
haloes by appealing to heating from radio-mode AGN feedback, following 
the accretion of material from the cooling flow onto a central supermassive 
black hole. 

{\it (3) Gas cooling: halo baryon fraction.} 
In Baugh2005M there is no heating of the hot halo by AGN 
feedback. Instead, a new channel is introduced for gas heated 
by the energy released by supernova explosions. Some fraction of the 
gas, as is common in the majority of semi-analytical models, is reheated 
and re-incorporated, on some timescale, into the hot gas halo (see Benson 
et~al. 2003). The rest of the reheated gas is ejected from the halo 
altogether in the superwind. In the Baugh2005 model, this gas is not allowed 
to re-cool at any stage. This process becomes inefficient in more massive 
haloes. However, the cooling rate is reduced in such haloes because they 
have less than the universal fraction of baryons (due to superwind ejection 
of gas from their progenitors). A detailed description of how superwinds are 
modelled in Baugh2005M is given in \citet{lacey.etal.2008}, who looked at
the properties of galaxies in the infra-red for comparison with observational
data from the \emph{Spitzer} space telescope.

{\it (4) Gas cooling: cooling in satellites.}
Font2008a introduced a new cooling scenario based on the hydrodynamical 
simulations of McCarthy et~al. (2008). Traditionally, the ram pressure 
stripping of the hot gas from a satellite halo has been assumed to be 
maximally efficient and instantaneous following a merger between 
two dark matter haloes. McCarthy et~al. showed that in gas simulations 
this is not the case and that the satellite can retain a substantial 
amount of hot gas, with the fraction depending upon the satellite orbit.
McCarthy et~al. used a suite of simulations to calibrate a recipe to 
describe how much hot gas is kept. Font et~al. (2008) extended the 
{\small GALFORM} code to include this prescription to calculate the amount 
of hot gas attached to each satellite galaxy within a halo and to allow 
the gas to cool directly onto the satellite, rather than onto the central 
(most massive) galaxy in the main dark matter halo. The other models
considered in this paper do not allow gas to cool onto satellite galaxies. 

{\it (5) Galaxy mergers.} 
Galaxies merge due to dynamical friction. 
Baugh2005M adopts the form of the merger timescale given by 
Eq.~4.16 of \citet{cole.etal.2000}. Bower2006a and Font2008a use 
the same prescription with a timescale that is longer by 50\%; 
physically this can be explained as a reduction in the mass of the 
satellite halo due to tidal stripping. DeLucia2006a use a hybrid scheme 
in which a satellite galaxy is associated with a substructure halo, 
which is followed until stripping and disruption result in it dropping 
below the resolution limit of the simulation. From the last 
radius at which the substructure was seen, an analytic estimate of the merger 
time is made, using the dynamical friction timescale (but with a different 
definition of the Coulomb logarithm) and applying a boost of a factor of 2 
(to improve the match to the bright end of the present day optical luminosity 
function). 

{\it (6) Star formation.} 
In the Durham models, the star formation timescale scales according to  
circular velocity of the disc computed at the half-mass radius. The 
star formation timescale also depends on a timescale parameter which can be 
held fixed (Baugh2005M) or which can scale with the disc dynamical time 
(Bower2006a, Font2008a). All of the cold gas is available for star formation. 
In the Munich model (DeLucia2006a), a critical mass of cold gas has to be 
reached before star formation can begin. This is motivated by the observational
inference that star formation requires a critical surface density of cold gas 
(which also has a theoretical motivation; cf. \citealt{1998ApJ...498..541K}).
Only the cold gas mass in excess of the threshold is available for star 
formation. The timescale adopted is the disc dynamical time. 
All the models adopt a standard solar neighbourhood initial mass function 
(IMF) for quiescent star formation in discs, although Baugh2005M 
adopts a top-heavy IMF with a correspondingly higher yield and recycled 
fraction for episodes of star formation triggered by galaxy mergers.

{\it (7) Heating of cooled gas by supernovae.} 
In the Durham models, the amount of gas reheated by supernova feedback per 
timestep is a multiple of the star formation rate, which depends principally
on the circular velocity of the disc as well as the choice of values adopted 
for the feedback parameters. As we have touched upon in (3) above, in 
Baugh2005M the gas reheated by supernovae can either be ejected completely in 
a superwind or heated up so that it is later re-incorporated into the hot halo 
(when a new halo forms i.e. after a doubling of the halo mass). 
Bower2006a and Font2008a do not consider the superwind channel for reheated 
gas. These models allow the reheated gas to be added to the hot gas reservoir 
on a timescale which depends on the halo dynamical time, rather than waiting 
for a new halo to form. DeLucia2006a follows \citet{croton.etal.2006}, who 
globally pins the rate at which gas is reheated by supernovae to a multiple of 
the star formation rate suggested by observations. The amount of energy 
released by supernovae is tracked and used to compute if any of the hot halo 
is ejected, to be re-incorporated on some timescale. \\ 

A common feature of all the models presented in this paper is
  that they contain parameters that are set by requiring that the predictions 
reproduce a subset of the available observational data. The primary 
consideration when setting the model parameters is that the model reproduces 
the present-day optical luminosity function as closely as possible. However, 
this alone is insufficient to set all of parameters, and so selected secondary 
observations are matched in order to specify the model. For example, the 
observed gas fractions in spirals and the sizes of discs are used to determine 
the Baugh2005M parameters, while Bower2006a focuses on reproducing the 
bimodality of the colour distribution of local galaxies
\citep[cf. ][]{gonzalez.etal.2008}. In both cases, constraining the parameters
in this way fixes the star formation timescales in the models. We refer the 
reader to the original references for a more complete discussion of which 
datasets are reproduced by the respective models. 

It is worth noting that, in the context of this study, gas fractions in 
spirals are the only data used to set parameters which explicitly relate to 
the cold gas content of galaxies\footnote{Even then, these data were used in 
only a subset of the models.}; other observations, such as the galaxy 
luminosity and mass functions, provide indirect constraints on the cold gas 
content. None of the model parameters have been adjusted for the purposes of 
this paper, except for the reduction in the cosmological baryon fraction in 
Baugh2005M, as explained above.

\section{Basic predictions}
\label{sec:results}

\begin{figure}
\includegraphics[width=8.5cm]{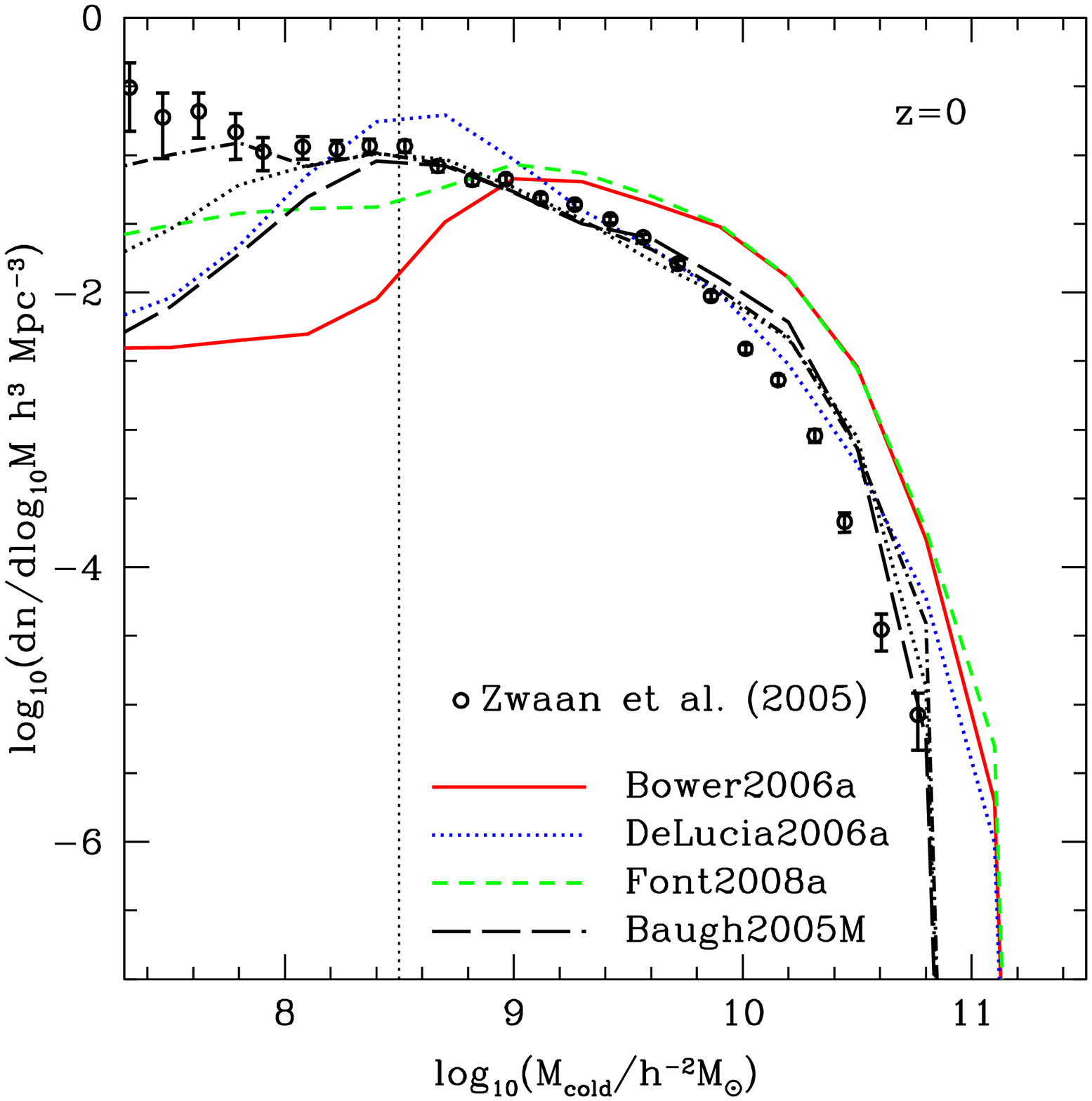}
\hfil
\includegraphics[width=8.5cm]{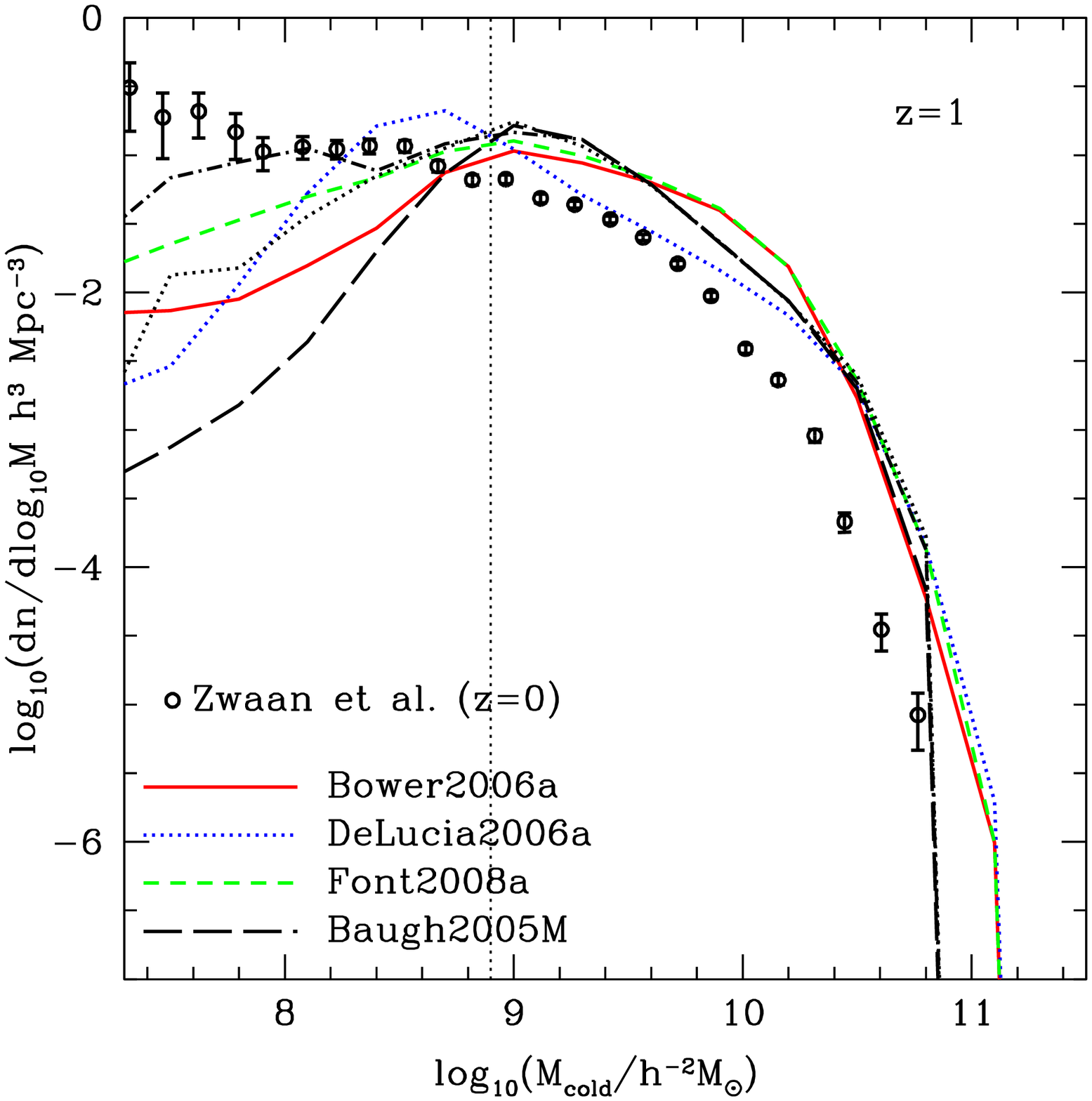}
\caption{The predicted cold gas mass function at $z=0$ (top) 
  and $z=1$ (bottom). The points show an observational 
  estimate of the HI mass function at z=0 by Zwaan et~al. (2005), converted 
  into a cold gas mass function by adopting the fixed H2/HI ratio 
    conversion factor described in \S\ref{ssec:convert_cg_to_hi}. 
  Different lines types correspond to different models as indicated 
  by the legend. For Baugh2005M, the results using 
  the Millennium simulation merger trees are shown by the dashed 
  black lines. The dotted and dot-dashed lines show calculations 
  using Monte Carlo merger trees with improved mass resolution (with 
  a mass resolution a factor of 2 and 4 better than the Millennium simulation
  respectively) but with the galaxy formation parameters held the same. 
  The dotted vertical lines indicate the cold gas mass resolution limit 
  of the Millennium galaxy formation models. The cold gas mass 
  resolution limit is slightly higher at $z=1$ than it is at $z=0$. In the 
  lower panel, the $z=0$ data points are repeated for reference. }
\label{fig:mcold}
\end{figure}

In this Section we present the model predictions for the cold gas masses, 
radii and rotation speeds of galactic discs. These quantities are used in the 
next section to predict the 21cm luminosity of the galaxies. Note that we do 
not discuss any quantities derived from these direct model outputs here, 
instead deferring such discussion until \S\ref{sec:observations}.\\

We begin by inspecting the cold gas mass functions predicted by the four 
models in Fig.~\ref{fig:mcold} at $z$=0 (upper panel) and $z$=1 (lower panel).
For comparison, we show also an ``observed'' $z$=0 cold gas mass function
(open circles and error bars), obtained by converting the $z$=0
mass function of HI in galaxies from HIPASS \citep[cf.][]{zwaan.etal.2005} 
to a cold gas mass function, using the ``fixed H2/HI ratio''
conversion factor discussed in \S\ref{ssec:convert_cg_to_hi}. The 
reader should note that cold gas masses are plotted in units of 
$h^{-2}\,\rm M_{\odot}$ rather than $h^{-1}\,\rm M_{\odot}$, which is the unit used
in simulations. This ensures that the observational units (which depend upon 
the square of the luminosity distance) are matched, but it introduces an 
explicit dependence on the dimensionless Hubble parameter $h$; here we adopt 
$h$=0.73, the value used in the Millennium simulation. 

We find that DeLucia2006a and Baugh2005M recover the observed $z$=0 cold gas 
mass function reasonably well, following the data closely between 
$M_{\rm cold} \simeq 10^{8.5} h^{-1}\,\rm M_{\odot}$ (approximately the cold gas mass 
resolution limit of the model; see below) and $M_{\rm cold} \simeq 10^{9.8} h^{-1} 
\, \rm M_{\odot})$; at larger $M_{\rm cold}$, both models tend to overestimate 
the amount of cold gas in galaxies by $\sim 0.25$ dex. In contrast, both 
Font2008a and Bower2006a predict systematically more cold gas in galaxies 
than is observed. This is unsurprising, however, because both Font2008a and 
Bower2006a also over-predict the gas-to-stellar ratio in spirals (see the 
discussion in \citealt{cole.etal.2000} and their figure 9). 

There is a minimum mass below which haloes are not reliably resolved in the 
Millennium simulation and this in turn imposes a minimum cold gas mass below 
which the predictions of the galaxy formation models are unreliable. This 
arises because the simulation can only recover the abundance of dark matter 
haloes down to some limiting mass\footnote{Typically this limiting mass is 
equivalent to $\sim 20$ particles, required for the halo mass function to be 
converged \citep[e.g.][]{jenkins.etal.2001}.}; below this limiting mass, the 
abundance of low-mass haloes will be suppressed because of finite mass 
resolution of the simulation. Furthermore, low-mass haloes may not be 
sufficiently well resolved for their merger trees to be considered reliable; 
this mass is likely to be larger than required for convergence of the halo 
mass function. 

This limiting halo mass is a problem because we expect cold gas to be present 
in haloes with masses below the resolution limit, and so we need to know how 
the limiting halo mass and the minimum cold gas mass relate to one another. 
This relationship can be estimated by running Monte Carlo merger trees of 
different minimum halo masses and comparing with the $N$-body merger trees. In 
practice, we run the Baugh2005M model with higher resolution trees generated 
using the new Monte Carlo prescription of \citet{parkinson.2008} and determine 
the halo mass down to which the Monte Carlo merger trees give a good match to 
the trees extracted from the Millennium simulation. The cold gas mass functions 
calculated using the $N$-body trees and the Monte Carlo trees diverge
below the mass indicated by the dotted vertical line in Fig. ~\ref{fig:mcold}, 
at a cold gas mass of $M_{\rm cold}= 10^{8.5} h^{-1}\,\rm M_{\odot}$. Note that we 
should repeat this exercise for each model in principle because the resolution 
limit may be sensitive to the model recipes. However, given the close agreement 
between the predictions above this mass limit, we do not expect the variation 
in the cold gas mass resolution between models to be large and so we expect the 
limiting mass obtained with Baugh2005M to be a reasonable estimate for all 
the models.\\

We note that there is little evolution in the predicted mass functions
back to $z=1$. This is remarkable because it shows that the sources
and sinks of cold gas more or less balance one another out. How can we 
understand this? We expect the sizes of galactic discs to decrease with 
increasing redshift. In three of the models (Bower2006a, 
DeLucia2006a and Font2008a) star formation proceeds on a timescale that is 
proportional to the 
circular orbit timescale in the disc, and so it follows that the star 
formation timescale decreases with increasing redshift. However, gas cools 
from the hot halo on a timescale that depends on local gas density; because 
density increases with increasing redshift, it follows that the cooling 
timescale also decreases with increasing redshift. Therefore we might expect 
that the amount of gas to cool per unit time will increase with increasing 
redshift but this is offset by the increasing numbers of stars that form per 
unit time with increasing redshift. This leads competing sources (gas
cooling) and sinks (star formation and mass ejection by winds) of cold
gas to balance each other. 

This explanation is compelling in its simplicity, but it is far from clear that 
it provides the complete picture of what is happening. Competition between 
sources and sinks of cold gas can plausibly balance each other, but it is worth
noting that there is little evidence for evolution of the cold gas mass 
function in Baugh2005M, which rests on assumptions that are quite 
different from those of Bower2006a, DeLucia2006a and Font2008a. In particular, 
the star formation timescale in Baugh2005M does not vary in proportion to the 
circular orbit timescale in the disc and so it is not obvious why the sources 
and sinks of cold gas should balance in this model, as they appear to. 
Understanding what physical processes drive the evolution of the cold gas 
mass function in the models is clearly an important issue, and one which shall
be the focus of future work.\\

\begin{figure}
\includegraphics[width=8.5cm]{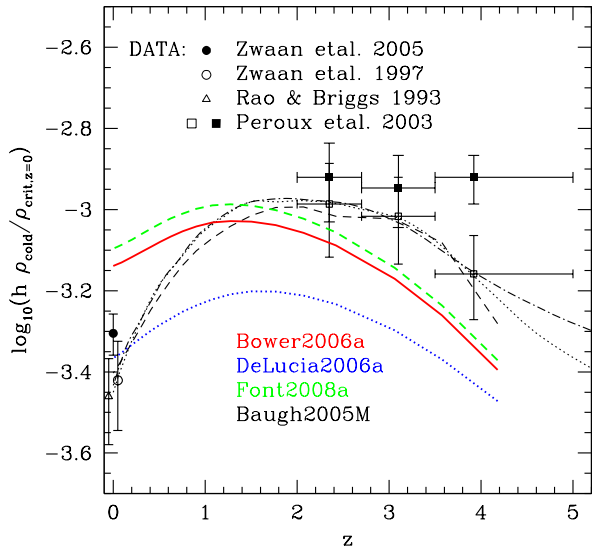}
\caption{
The predicted cold gas density $\rho_{\rm cold}$, normalised by the 
value of the critical density $\rho_{\rm crit}$ at $z$=0, as a function of $z$. 
Different lines correspond to different models as indicated by the 
legend. For Baugh2005M, the results using the Millennium 
simulation merger trees are shown by the dashed black line. The dotted 
and dot-dashed lines show calculations using Monte Carlo merger trees 
with improved mass resolution but with the galaxy formation parameters 
held the same (see caption of Fig.~\ref{fig:mcold}). 
Filled and open circles correspond to \citet{zwaan.etal.1997,zwaan.etal.2005}
data respectively; open triangles correspond to \citet{rao.and.briggs.1993}; 
open and filled squares correspond to \citet{peroux.etal.2003}. In the 
latter case, the open squares indicate the cold gas density inferred from 
damped Lyman $\alpha$ systems by Peroux et~al., and the filled squares include
a correction to take into account gas clouds with lower column density.} 
\label{fig:global_HI}
\end{figure}

We now consider the global mass density of cold gas 
$\Omega_{\rm cold}=\rho_{\rm cold}/\rho_{\rm crit}$. In Fig.~\ref{fig:global_HI} we 
show how $\rho_{\rm cold}$ varies with redshift $z$, normalised by 
$\rho_{\rm crit}/h$ at $z$=0 for ease of comparison with observational data.
This reveals that both Bower2006a and Font2008a over-predict the density of 
cold gas at $z <1$ and somewhat under-predict the amount of cold gas at higher 
redshifts. DeLucia2006a predicts a cold gas density that is consistent with 
observational estimates at $z$=0 but it under-predicts the density at $z>0$ by 
a factor of two to three. Of all the models, Baugh2005M most closely matches 
the observed density of cold gas at all redshifts. Comparing the predictions 
for this model using the Millennium simulation merger trees with those from 
Monte Carlo merger trees (with improved mass resolution) suggests that the 
$N$-body results are robust up to $z \sim 4$. \\

\begin{figure}
\includegraphics[width=8.5cm]{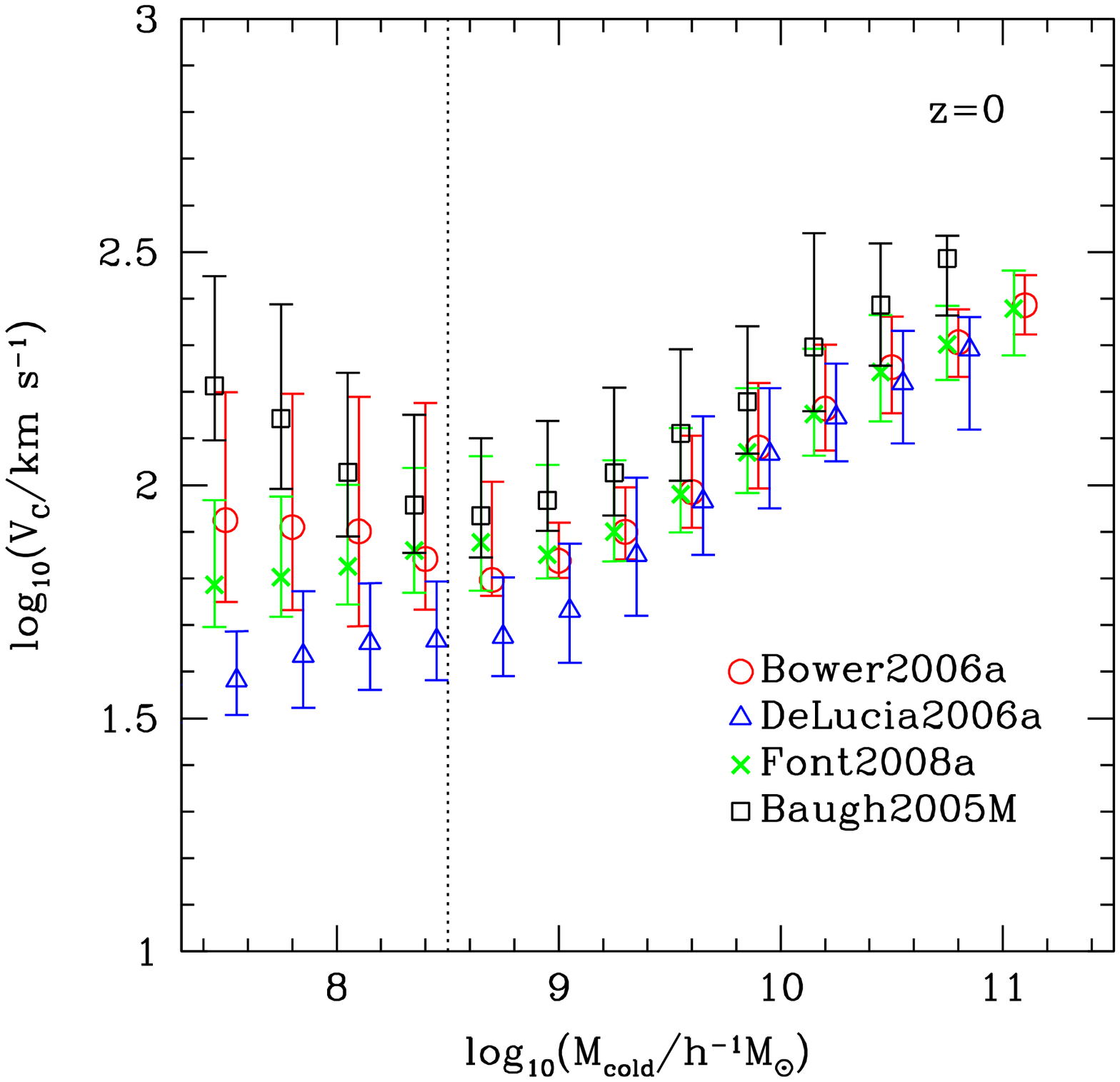}
\includegraphics[width=8.5cm]{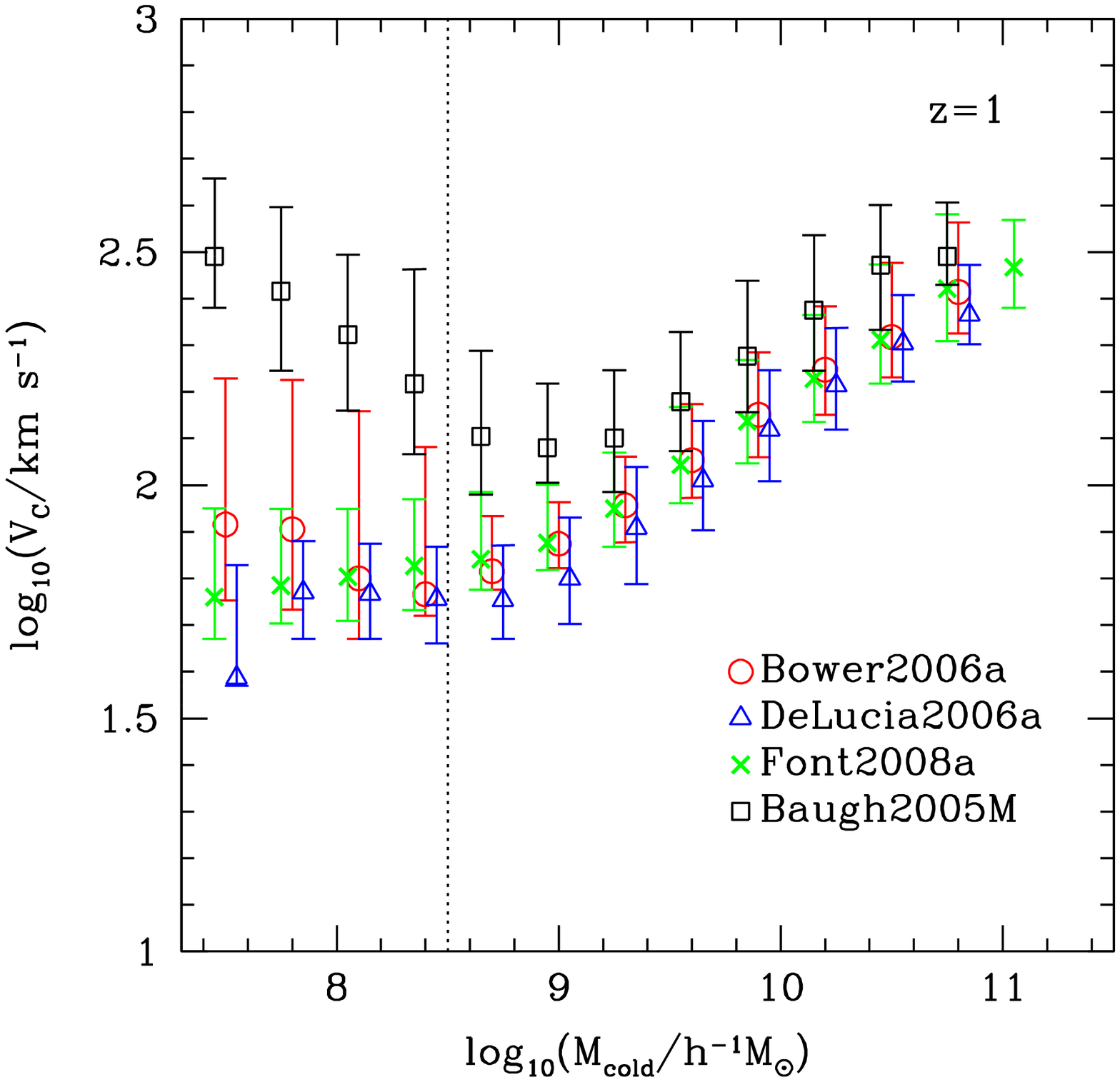}
\caption{The predicted circular velocity - cold gas mass relation 
at $z=0$ (top) and $z=1$ (bottom). The points shown the median 
velocity and the bars show the 10-90 percentile range. Different 
symbols correspond to different models as indicated by the legend. 
In DeLucia2006, the velocity plotted is measured at the virial radius 
of the dark matter halo; in the other cases, it is the circular velocity 
at the half-mass radius of the disc. The dotted vertical line indicates 
the cold gas mass resolution limit of the Millennium galaxy formation models. 
}
\label{fig:vdisc}
\end{figure}

\begin{figure}
\includegraphics[width=8.5cm]{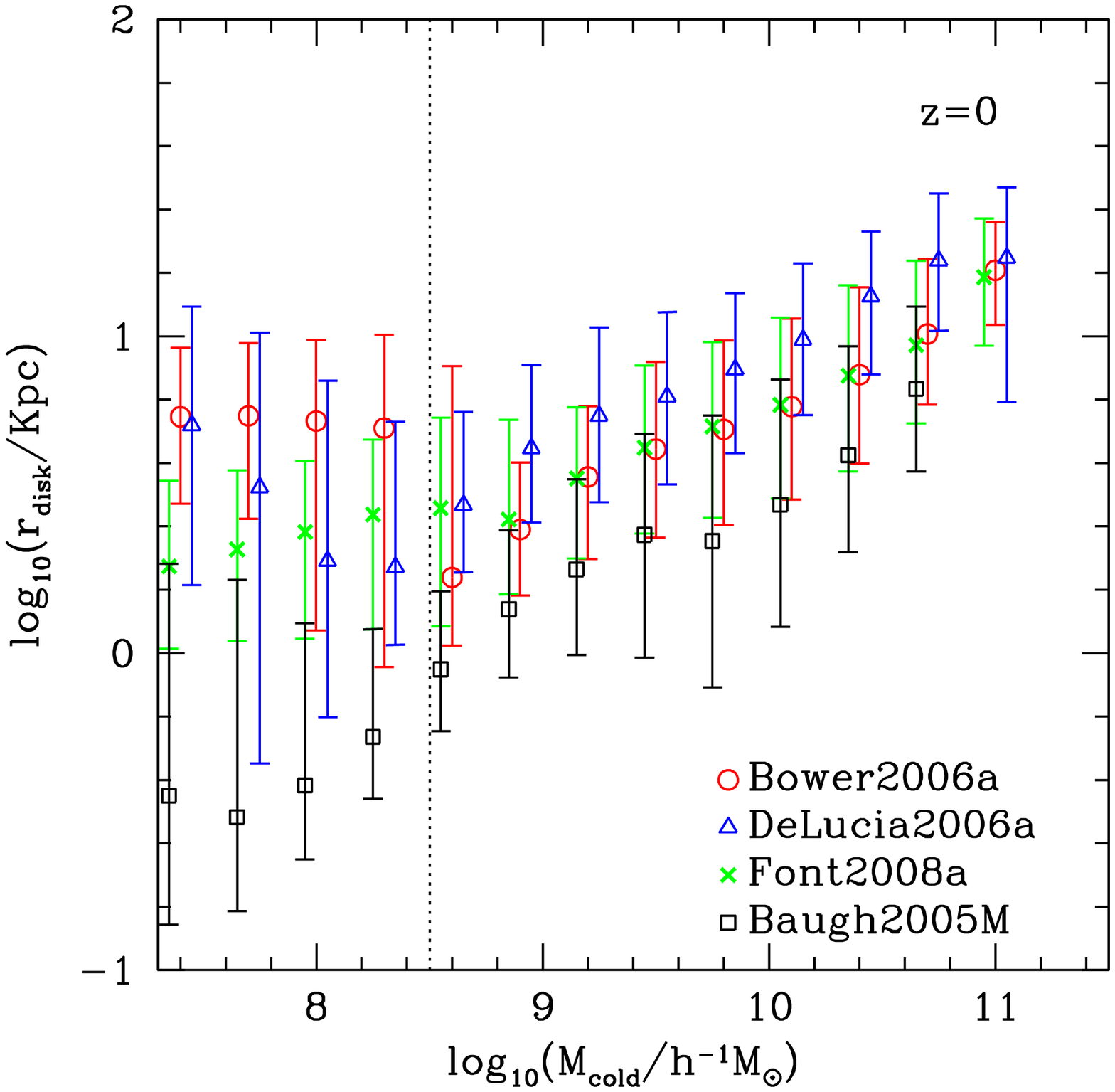}
\includegraphics[width=8.5cm]{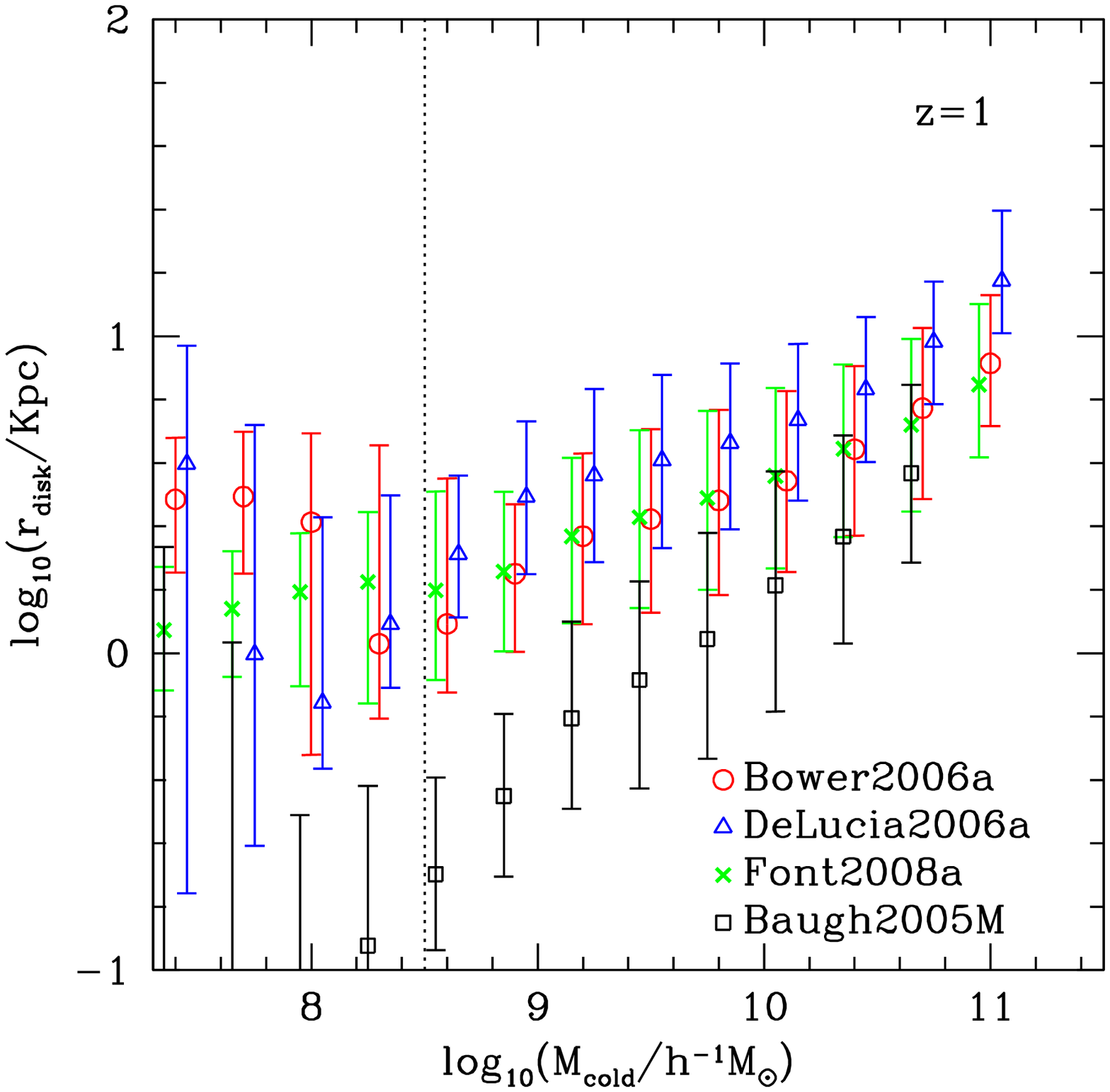}
\caption{The predicted disc radius - cold gas mass relation 
at $z=0$ (top) and $z=1$ (bottom). The points show the median 
velocity and the bars show the 10$^{\rm th}$-90$^{\rm th}$ percentile range. 
Different symbols correspond to different models as indicated by the legend. 
The radius plotted is the half-mass radius of the disc. In DeLucia2006, 
the quantity stored in the Millennium Archive is three times the 
scale-length of the exponential disc, which we convert to a half-mass radius. 
The dotted vertical line indicates the cold gas mass resolution limit 
of the Millennium galaxy formation models. }
\label{fig:rdisc}
\end{figure}

We now compare the rotation speeds of galactic discs as a function of 
cold gas mass. This is interesting to quantify because it indicates how the
velocity width is likely to scale with HI mass, which is important for HI
surveys. It also provides a useful insight into how the mean cold gas mass
varies as a function of galaxy mass, which can be related to the rotation 
speed. 

Fig.~\ref{fig:vdisc} shows the rotation speed - cold gas mass relation for 
galactic discs at $z=0$ (top) and $z=1$ (bottom) predicted by the models. 
Note that the Durham and Munich models define rotation speed in different ways;
in the Durham models, the rotation speed plotted is the circular velocity at 
the half-mass radius of the disc, whereas in the Munich model (DeLucia2006a), 
the rotation speed plotted is the circular velocity measured at the virial 
radius of the dark matter halo. The precise relationship between the circular 
velocities measured at the half-mass radius of the disc and at the virial 
radius of the host halo depends on the mass and distribution of the cold gas 
and stars in the disc and bulge and of the dark matter. For example, in 
Bower2006a, we find that the circular velocity
at the half-mass radius of the disc is typically 20\% higher than that 
measured at the virial radius of the host halo, for $L^{\ast}$ galaxies. After 
allowing for this difference, the DeLucia2006a rotation speed-cold gas mass 
relation is in close agreement with the predictions of
Bower2006a and Font2008a. This level of agreement is quite remarkable 
given the differences in the implementations of the physical ingredients in 
the models. In contrast, Baugh2005M predicts a rotation speed that is higher 
than the other models by around 50\%. One possible explanation for this is 
that discs are more compact in this model, which is indeed the case in 
Fig.~\ref{fig:rdisc} (see below). 

The model predictions diverge from each other below a cold gas mass of 
$M_{\rm cold}= 10^{8.5} h^{-1}\,\rm M_{\odot}$ and there is a change in the slope 
of the rotation speed - cold gas mass relation below this mass. This is the 
minimum mass down to which the predictions from the Millennium simulation 
merger trees are reliable. There is very little evolution in the rotation 
speed - cold gas mass relation between $z=0$ and $z=1$; the zero-point of 
the $z=1$ relation is about 25\% higher.\\

The predicted disc radius - cold gas mass relation is plotted in 
Fig.~\ref{fig:rdisc}. For the Durham models, the disc radius plotted 
corresponds to the half-mass radius of the disc, which is calculated by taking
into account conservation of the angular momentum of the cooling gas and 
the dynamical equilibrium of the disc, bulge and dark matter \footnote{See 
\citet{cole.etal.2000}, \citet{almeida.etal.2007} and 
\citet{gonzalez.etal.2008} for details of this calculation.}.
For the Munich model (DeLucia2006a), the quantity stored in the Millennium 
Archive is three times the scale-length of the exponential disc, which is 
computed by scaling from the virial radius of the dark matter halo. We convert
this length to a half-mass disc radius to plot on Fig.~\ref{fig:rdisc}. 

The DeLucia2006a half-mass radii estimated in this way are approximately 
$0.2-0.3$ dex larger than those predicted by Bower2006a and Font2008a. 
In contrast, Baugh2005M predicts smaller gas discs than the other models, but 
we note that this model also predicts sizes for stellar discs that are in much 
better agreement with observational data at $z=0$ than the other models 
\citep{gonzalez.etal.2008}. As with the rotation speed - cold gas mass 
relation, there is little evolution in the disc radius - cold gas mass 
relation between $z=0$ and $z=1$. The model predictions diverge from one 
another below the mass resolution of $M_{\rm cold} = 10^{8.5} h^{-1}\,\rm 
M_{\odot}$. 

It is worth remarking that it is surprising that the DeLucia2006a, Bower2006a 
and Font2008a predictions are so close, given the differences in the way the 
disc sizes are computed in the models. One might expect the half-mass radii 
predicted by the Durham models to be smaller than those from the Munich model 
because the former take into account the gravitational contraction of the dark 
matter and the self-gravity of the disc and bulge, whereas the latter adopts a 
simple scaling of the virial radius of the host halo.

\section{Implications for HI Surveys}
\label{sec:observations}

The results presented so far encapsulate what current semi-analytical galaxy 
formation models can tell us about the distribution of cold gas masses of 
galaxies as a function of redshift. We can use this information to deduce the
distribution of HI masses of galaxies, from which we can predict
number counts of HI sources as a function of redshift. These predictions 
can then be compared with forthcoming HI surveys on the SKA
pathfinders, such as ASKAP, MeerKAT and APERTIF, and ultimately on the SKA. 

Such a comparison represents an important and fundamental test of the
semi-analytical galaxy formation framework. Semi-analytical model parameters are
calibrated explicitly to reproduce statistical properties of the observed
galaxy population where observational data exist, such as the galaxy 
luminosity function \citep[e.g.][]{benson.etal.2003.b} and the abundance 
of sub-mm galaxies at high redshift \citep[e.g.][]{baugh.etal.2005}. 
However, this approach is sometimes criticised precisely because it is 
calibrated to reproduce properties of the observed galaxy population. It 
is not always clear how robust model predictions are if the parameters have 
been adjusted to match as many observational datasets as possible. 
Few observed data exist for the HI properties of galaxies at 
redshifts $z \gtrsim 0.05$ and so such data will provide a compelling 
test of the currently favoured models we consider in this paper.

In this section we use the cold gas mass functions presented in the previous
section to predict HI source number counts for forthcoming HI surveys. To do
this, we must first convert cold gas masses, which are the natural outputs of 
the models, to HI masses. Then we consider how the sensitivity and
angular resolution of a radio telescope affects whether or not a particular 
galaxy at a given redshift is likely be detected in a given HI survey. Finally
we investigate the impact of sensitivity on the number counts of HI galaxies
in ``peak flux limited'' surveys and we assess the angular resolution required 
to resolve gas-rich galaxies out to $z \sim 3$. 

\subsection{Conversion of Cold Gas Mass to HI Mass}
\label{ssec:convert_cg_to_hi}

The results presented in \S\ref{sec:results} are for cold gas 
  masses in galaxies, but we require HI masses. How should we convert from cold 
  gas to HI mass?

\begin{itemize}
\item First, we note that $ \sim 24\%$ by mass of this cold gas will be in 
  the form of helium; this leaves $\sim 76\%$ by mass in the form of hydrogen. 
\item Second, we note that this $\sim 76\%$ hydrogen will be split into neutral 
  (atomic, molecular) and ionised fractions, but for simplicity we assume that 
  the ionised fraction in the disc is sufficiently small that we can ignore it.
\item Third, we must determine what fraction of the neutral hydrogen is 
  molecular in form; this then allows us to assign an HI mass to each galaxy,
  given its cold gas mass.
\end{itemize}

It is worth providing some justification for our argument that the ionised
fraction is small. Recall that we consider cold gas to be gas that has cooled 
radiatively from a hot phase to below $10^{4}\rm K$ and is available for star 
formation (cf. \S\ref{sec:galform}). At a temperature of $\lesssim 10^4 
\rm\,K$, this cold gas will include warm hydrogen in its atomic and ionised 
phases \citep[cf.][]{ferriere.2001}. Observations tell us that the ratio
of ionised to atomic hydrogen in the 
midplane of the Galaxy is small ($\sim 5\%$) but it increases with increasing 
scale-height, and by scale-heights $\gtrsim 1$ kpc the warm ionised state 
probably dominates \citep[cf.][]{reynolds.2004}. The typical 
(i.e. full width at half maximum) scale-heights of atomic and molecular 
hydrogen are much smaller than this ($\sim 100-200$pc), and so what one
estimates for the ionised mass within the disc depends on the range of 
scale-heights included. We adopt the mass estimates of \citet{ferriere.2001}
for the total ionised and neutral hydrogen masses of the Galaxy 
($\gtrsim 7.5 \times 10^9 \rm M_{\odot}$ and $\gtrsim 1.6 \times 10^9 \rm 
M_{\odot}$) to estimate that the ionised fraction constitutes approximately
$15\%$ by mass of the Galactic disc. This is sufficiently small that we can 
ignore it for the purposes of this study, although more detailed modelling would
need to take it into account.

Of the remaining $\sim 76\%$ by mass of cold gas that is in the form of 
neutral hydrogen, what is the ratio of molecular ($\rm H_2$) to atomic
(HI) hydrogen? We consider two approaches;

\begin{itemize}

\item A ``fixed $\rm H_2/HI$'' ratio for all galaxies for all redshifts
  \citep[cf.][]{baugh.etal.2004}.

\item A ``variable $\rm H_2/HI$'' ratio that depends on galaxy properties 
  and redshift \citep[cf.][]{obreschkow2009a,obreschkow2009c}.

\end{itemize}

\noindent The fixed $\rm H_2/HI$ approach was used in the 
\citet{baugh.etal.2004} study and it allows us to apply a simple uniform 
scaling to the cold gas mass functions presented in \S\ref{sec:results} 
to obtain HI mass functions. It is a purely empirical scaling in the sense 
that it uses estimates of the global $\rm H_2$ and $\rm HI$ densities in 
the local Universe to deduce the ratio of molecular to atomic hydrogen. 
\citet{baugh.etal.2004} used the estimates of \citet{keres.etal.2003} and 
\citet{zwaan.etal.2005} respectively for the global $\rm H_2$ and $\rm HI$ 
densities 
($\rho_{\rm H_2} = (3.1 \pm 0.9) \times 10^7 h\,\rm M_{\odot} \rm Mpc^{-3}$
and $\rho_{\rm HI} = (8.1 \pm 1.3) \times 10^7 h \, \rm M_{\odot} \rm Mpc^{-3}$)
to deduce a ratio of molecular to atomic hydrogen of $\sim 0.4$. This 
gives a conversion factor of
\begin{equation}
  \label{eq:fixed}   
  M_{\rm HI} = 0.76\,M_{\rm cold}/(1+0.4) \simeq 0.54\,M_{\rm cold},
\end{equation}
\noindent which is the one we adopt. 

The variable $\rm H_2/HI$ approach is based on the work of 
\citet{blitz.2006}, \citet{leroy.etal.2008} and \citet{obreschkow2009b},  
and it allows us to estimate the $\rm H_2/HI$ ratio on a galaxy-by-galaxy 
basis. There have been various attempts to predict theoretically the 
variation of the $\rm H_2/HI$ ratio on galactic scales based on physical 
models of the ISM; for example, \citet{elmegreen.1993} argued that the 
most important physical parameters driving variations in the $\rm H_2/HI$ 
ratio are the gas pressure and the local intensity of the interstellar UV
radiation field, while \citet{krumholz.etal.2009} argued instead that
the main physical parameters driving variations are the column density
and metallicity of interstellar gas clouds. On the other hand, \citet{wong2002}
found from spatially resolved observations of nearby galaxies that the ratio 
of $\rm H_2$ to HI surface densities increases with increasing midplane 
hydrostatic gas pressure, following a power-law relation 
$\Sigma_{\rm H_2}/\Sigma_{\rm HI} \propto P_0^{\alpha}$. This empirical 
power-law relation was then confirmed in more detailed observational 
studies by \citet{blitz.2006} and \citet{leroy.etal.2008}, and found to 
extend from $\Sigma_{\rm H_2}/\Sigma_{\rm HI} \ll 1$ to 
$\Sigma_{\rm H_2}/\Sigma_{\rm HI} \gg 1$. Note that the midplane pressure 
$P_0$ used in these relations is not directly measured, but is instead 
inferred from the gas and stellar surface densities combined with assumed 
velocity dispersions or scale-heights, assuming hydrostatic equilibrium.

Building on this work, \citet{obreschkow2009b} have derived a model for the 
global $\rm H_2/HI$ ratio in a galaxy. This uses the 
$\Sigma_{\rm H_2}/\Sigma_{\rm HI} = (P_0/P_{\ast})^{\alpha}$ relation in
the form found by \citet{leroy.etal.2008}, with $\alpha=0.8$ and
$P_{\ast}=2.35\times 10^{-13} {\rm Pa}$, and assumes that the stars and gas 
in a galactic disc both have exponential profiles, though with different 
radial scale-lengths. After setting the free parameters of the model by 
comparison with observational data on nearby galaxies, 
\citeauthor{obreschkow2009b} obtain the following expression for the global 
$\rm H_2/HI$ ratio, $R_{\rm mol}^{\rm gal}=M_{\rm H_2}/M_{\rm HI}$:

\begin{equation}
  \label{eq:rmolgal}
  R_{\rm mol}^{\rm gal}=\left(3.44\,{R_{\rm mol}^c}^{-0.506} + 4.82\,{R_{\rm mol}^c}^{-1.054} \right)^{-1},
\end{equation}
\noindent where $R_{\rm mol}^c$ is the $\rm H_2/HI$ ratio at the centre of the disk, given by
\begin{equation}
  \label{eq:rmolc}
  R_{\rm mol}^c=\left[K\,r_{\rm disc}^{-4}\,M_{\rm gas}\left(M_{\rm gas}+
    \left<f_{\sigma}\right>\,M_{\rm disc}^{\ast}\right)\right]^{0.8}.
\end{equation}
\noindent In the above expressions, $M_{\rm disc}^{\ast}$ and $M_{\rm gas}$ are 
the masses of stars and gas in the disc, and $r_{\rm disc}$ is the exponential 
scale-length of the gas, while $K=G/(8\pi\,P_{\ast})=11.3\,\rm m^4\,kg^{-2}$, and
$\left<f_{\sigma}\right>=0.4$ is the average ratio of the vertical velocity 
dispersions of gas to stars. Given $R_{\rm mol}^{\rm gal}$, the conversion factor 
between cold gas mass and HI mass is then
\begin{equation}
  \label{eq:variable}  
M_{\rm HI} = 0.76\,M_{\rm cold}/(1+R_{\rm mol}^{\rm gal}).
\end{equation}
We employ Eq.~\ref{eq:rmolgal}, Eq.~\ref{eq:rmolc} and Eq.~\ref{eq:variable} 
using the values of $M_{\rm disc}^{\ast}$ and $M_{\rm gas}$ predicted by the 
semi-analytical models. Because neither the Durham nor Munich models 
distinguish between the half-mass radii of the stars and gas, we take the 
gas half-mass radius to be equal to the total half-mass radius predicted by 
the semi-analytical model, and we then convert it to a disc scale-length
$r_{\rm disc}$ by assuming an exponential disc (as in 
\citeauthor{obreschkow2009b}). 

\begin{figure}
  \centering
  \includegraphics[width=8.0cm]{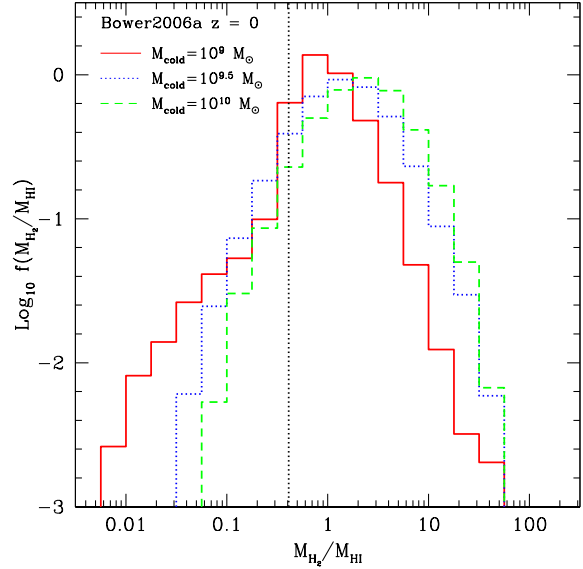}
  \caption{Distribution of the ratio $M_{\rm H_2}/M_{\rm HI}$
      ($R_{\rm mol}^{\rm gal}$) predicted by the variable $\rm
      H_2/HI$ approach for galaxies with cold gas masses $\rm M_{\rm
      cold}$=$10^9/10^{9.5}/10^{10} \,\rm M_{\odot}$ (solid, dotted and dashed
    histograms respectively) in Bower2006a at $z$=0. We include all galaxies
    within a bin of 0.3 dex centred on $\rm M_{\rm cold}$. The histograms are
    normalised by the area under the curve. The light dotted vertical line 
    indicates the ratio assumed if the fixed $\rm H_2/HI$ approach is
    used.}
  \label{fig:comp_mcold_mhi}
\end{figure}

How significant is the difference between the HI masses estimated
assuming a variable $\rm H_2/HI$ approach and those estimated assuming a 
fixed $\rm H_2/HI$ approach? Fig.~\ref{fig:comp_mcold_mhi} shows the 
distribution of the ratio $M_{\rm H_2}/M_{\rm HI}$ (i.e. $R_{\rm mol}^{\rm gal}$) for 
galaxies at $z$=0 in Bower2006a. This is interesting because 
$R_{\rm mol}^{\rm gal}$ fixes $M_{\rm HI}$ for a galaxy, given $M_{\rm cold}$ 
(cf. Eq.~\ref{eq:variable}), 
and so knowledge of the distribution of $R_{\rm mol}^{\rm gal}$ provides important
information about the distribution of $M_{\rm HI}$. We split our galaxy sample by
$M_{\rm cold} $ into three mass bins, of width 0.3 dex, centred on 
$M_{\rm cold} $=$10^9$,$10^{9.5}$ and $10^{10} \,\rm M_{\odot}$ (solid, dotted 
and dashed histograms); these contain 55042, 60579 and 32213 galaxies 
respectively. For each mass bin, we construct a histogram of 
$M_{\rm H_2}/M_{\rm HI}$ estimated using Eq.~\ref{eq:rmolgal} and we
normalise it by the area under the curve. For comparison, we indicate
also the ratio $M_{\rm H_2}/M_{\rm HI}$ one obtains assuming a 
fixed $\rm H_2/HI$ approach by the light dotted vertical line.

Fig.~\ref{fig:comp_mcold_mhi} is striking because it shows that 
variable $\rm H_2/HI$ approach predicts a broad distribution of 
$M_{\rm H_2}/M_{\rm HI}$ in each mass bin. The medians of the distributions are
$\sim 0.9$, $\sim 1.5$ and $\sim 2.2$ in the bins centred on $M_{\rm cold} $=
($10^9$/$10^{9.5}$/$10^{10}$) $\rm M_{\odot})$, compared to $\sim 0.4$ assumed 
in the fixed $\rm H_2/HI$ approach; indeed, only $\sim 11\%$ of the 
galaxies in the $M_{\rm cold} $=$10^9$/$10^{9.5}\,\rm M_{\odot}$ mass bins and 
$\sim 5\%$ in the $M_{\rm cold} $=$10^9\,\rm M_{\odot}$ mass bin have 
$M_{\rm H_2}/M_{\rm HI}$ ratios as small as this. Physically, this means that
a typical galaxy will have a smaller fraction of its cold gas mass in the form
of HI  -- by as much as a factor of $\sim 100$ -- in the variable 
$\rm H_2/HI$ approach than in the fixed $\rm H_2/HI$ approach. 
Inspection of Eq.~\ref{eq:rmolc} suggests that this behaviour reflects the
strong scaling with $r_{\rm disc}$ ($\propto r_{\rm disc}^4$). Within a given mass
bin there is a distribution of $r_{\rm disc}$ and the $10^{\rm th}$ and 
$90^{\rm th}$ percentiles of this distribution differ by a factor of at least
a few with respect to the median, translating into variations of factors of
$\sim 20-100$ in surface density and consequently local gas pressure with
respect to the median. This implies that the $M_{\rm H_2}/M_{\rm HI}$ ratio can 
in principle vary by factors of $\sim 10^2$ to $10^4$ within a given mass bin, 
which will affect the number of HI sources that can be detected. 

\begin{figure}
  \centering
  \includegraphics[width=8.0cm]{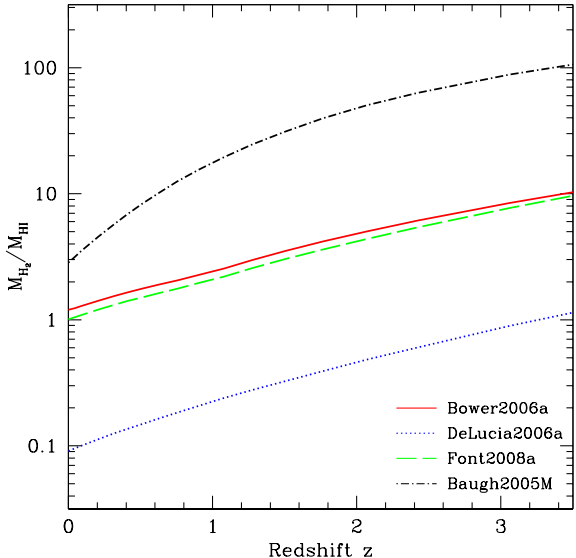}
  \caption{Variation of the median $M_{\rm H_2}/M_{\rm HI}$ 
      (i.e. $R_{\rm mol}^{\rm gal}$) of all galaxies with $M_{\rm cold} \geq 
      10^{8.5}\,\rm M_{\odot}$ with redshift. Different lines correspond 
      to different models as indicated by the legend.}
  \label{fig:comp_h2h1_z}
\end{figure}

The dependence on $r_{\rm disc}$ in the variable $\rm H_2/HI$ approach 
implies that the mean/median $M_{\rm H_2}/M_{\rm HI}$ ratio should increase 
sharply with increasing redshift, which in turn implies that HI masses of 
galaxies should be smaller at high redshifts. This is made clear in 
Fig.~\ref{fig:comp_h2h1_z}, which shows how the median of the distribution 
of $M_{\rm H_2}/M_{\rm HI}$ for all galaxies with 
$M_{\rm cold} \geq 10^{8.5}\,\rm M_{\odot}$ varies with redshift for the four 
models. In all cases, the median increases with redshift, as we would expect. 
However, the most striking aspect of this figure is the pronounced offsets 
between different models. The DeLucia2006a medians are an order of magnitude 
smaller than the Bower2006a/Font2008a medians at all redshifts. In contrast, 
Baugh2005M predicts a median that is a factor of a few larger at $z$=0 than 
the Bower2006a/Font2008a medians, but the difference grows to a factor of 
$\sim 10$ by $z \sim 3$. The models predict broadly similar cold gas and 
stellar masses and so it is the differences in scale-lengths, apparent in 
Fig.~\ref{fig:rdisc}, and the strong scaling with disc scale-length 
($\propto r_{\rm disc}^4$) that drive these large offsets. The net effect of this 
strong variation of $M_{\rm H_2}/M_{\rm HI}$ with redshift will be to dramatically
reduce the number of HI sources detectable at higher redshift (see 
Fig.~\ref{fig:dndz_variable} in the next section). 

This demonstrates that how one chooses to calculate a galaxy's HI mass is 
important. However, for the purposes of this study, we adopt the 
fixed $\rm H_2/HI$ approach to converting cold gas masses to HI masses. The 
variable $\rm H_2/HI$ approach has been calibrated using observations of
galaxies in the local Universe. There are sound physical reasons to expect
that there will be a correlation between local gas pressure and molecular
fraction in galactic discs at all redshifts, but it is unclear how reliable 
the local correlation is likely to be when applied to high redshifts. In
contrast, the fixed $\rm H_2/HI$ approach provides a reasonable upper
bound to the number of sources we might expect to detect. 

\subsection{Detection of HI Sources}

Two issues are key in determining whether or not an HI source will be
detected reliably by a radio telescope or interferometer; namely,
sensitivity and angular resolution. An HI source has an intrinsic
21-cm luminosity that depends primarily on its HI mass $M_{\rm HI}$ which,
along with its line-of-sight velocity width $\Delta V_{\rm los}$ and distance 
$D$,
determines the flux at the position of the observer. The observer
measures this flux with a receiver that has finite sensitivity
determined primarily by its effective collecting area $A_{\rm eff}$
and the system temperature $T_{\rm sys}$, and it is this limiting
sensitivity that determines whether or not the source is
detected. Note, however, that angular resolution also plays an
important role; HI 21-cm emission from a galaxy is likely to be
spatially extended and an extended source can be ``resolved out'' by
an interferometer if it is observed with too high an angular resolution. 
In the next two sections we consider how sensitivity and angular 
resolution affect HI source number counts.

\subsubsection{Sensitivity}
\label{ssec:sensitivity}

If we could construct the ideal radio telescope with arbitrarily high
sensitivity, then we would observe a flux $S_{\rm obs}$ from an HI source
at redshift $z$. This is determined by the source's HI mass $M_{\rm HI}$, its
velocity width $\Delta V_{\rm los}$ and redshift $z$. The relationship between 
$S_{\rm obs}$ and $M_{\rm HI}$, $\Delta V_{\rm los}$ and $z$ can be obtained as 
follows.

The emissivity $\epsilon_{\nu_0}$ at rest-frame frequency $\nu_0$ tells us the rate at which 
energy is emitted by an HI source at this frequency per unit volume per  
steradian. We can express this as

\begin{equation}
  \label{eq:emissivity}
  \epsilon_{\nu_0} = \frac{1}{4\pi} h \nu_0 A_{12}  \frac{n_2}{n_{\rm H}} n_{\rm H} \phi(\nu_0),
\end{equation}

\noindent where $h\nu_0$ is the energy of the 21-cm photon ($h$ is Planck's
constant and $\nu_0$ is the photon frequency), $n_{2}/n_{\rm H}$ tells us 
what fraction of atoms are expected to be in the upper state, $A_{12}$ is the 
Einstein coefficient which tells us the spontaneous rate 
of the transition from the upper to lower state, $n_{\rm H}$ is the total 
number density of hydrogen atoms in the source and $\phi(\nu)$ is the 
line profile. We expect $n_2/n_{\rm H} \simeq 3/4$ because the 
temperature of the cloud corresponds to a much larger energy than the 
energy difference corresponding to the transition from the upper to lower 
state \citep[i.e. $kT \gg h \nu$, cf.][]{spitzer1978}. Integrating over a 
solid angle of $4\pi$ steradians and over the volume of the source 
gives us the luminosity at frequency $\nu_0$, $L_{\nu_0}$,

\begin{equation}
\label{eq:luminosity}
L_{\nu_0} = \frac{3}{4} h \nu_0 A_{12} \frac{M_{\rm HI}}{m_{\rm H}} \phi(\nu_0),
\end{equation}

\noindent where we write the number of hydrogen atoms as $M_{\rm HI}/m_{\rm H}$, 
$m_{\rm H}$ being the mass of the hydrogen atom. 

When we observe 21-cm emission from an HI source, the radiation arises
from a forbidden transition, which implies a small natural line width
($5 \times 10^{16}$ Hz). Therefore the observed line profile
$\phi(\nu_0)$ is in practice determined by Doppler broadening due to
the motions of HI atoms in the galaxy, which, in disc galaxies, are dominated
by the large-scale rotational velocity. We therefore assume
that $\phi(\nu_0)$ can be approximated as a top hat function of width
$\Delta\nu_0 = (\Delta V_{\rm los}/c) \nu_0$ and height $1/\Delta\nu_0$.
Noting this, we can write the total monochromatic flux at the position of 
the observer as

\begin{equation}
\label{eq:source_flux}
S_{\nu} = (1+z) \frac{L_{\nu (1+z)}}{4\pi D_{\rm L}(z)^2} ;
\end{equation}

\noindent here $\nu=\nu_0\,(1+z)^{-1}$ is the redshifted frequency 
measured by the observer and $D_{\rm L}(z)=(1+z)\,D_{\rm co}(z)$ is the 
luminosity distance of the source with respect to the observer ($D_{\rm co}(z)$
is the radial comoving separation between source and observer). Therefore the 
measured flux at the position of the observer is 

\begin{equation}
  \label{eq:source_flux_mass}
  S_{\rm obs}\,\Delta \nu= \frac{3}{16 \pi} \frac{h \nu A_{12}}{m_{\rm H}} M_{\rm HI} \frac{1}{D_L(z)^2}(1+z),
\end{equation}

\noindent which we rewrite as

\begin{equation}
  \label{eq:source_flux_mass2}
  S_{\rm obs}= \frac{3}{16 \pi} \frac{hcA_{12}}{m_{\rm H}} M_{\rm HI}
  \frac{1}{D_L(z)^2}\frac{1}{\Delta V_{\rm los}} (1+z).
\end{equation}

\noindent Here we assume that $\Delta\nu$ in Eq.~\ref{eq:source_flux_mass} 
can be written as 

\begin{equation}
\label{eq:deltanu}
  \Delta\nu = \frac{\Delta\nu_0}{1+z} = \frac{\Delta V_{\rm los}}{c}\,\frac{\nu_0}{1+z},
\end{equation}

\noindent where $\Delta V_{\rm los}$ is the rest-frame line-of-sight velocity 
width of the galaxy. $\Delta V_{\rm los}$ will depend on the inclination $i$ of 
the disc, varying as $2\,V_{\rm c}\,\sin i$ where $V_{\rm c}$ is the disc circular
velocity, and $i=0$ or $\pi/2$ correspond to a face-on or edge-on disc 
respectively. In our analysis we assume that galaxy discs have random 
inclinations with respect to the observer, with an average velocity width of 
$\sim\!1.57\, V_{\rm c}$ (cf. \S\ref{ssec:observables}).\\

The measured flux from the source must be compared with the 
intrinsic limiting sensitivity of the receiver. Assuming a dual polarisation
radio receiver, the limiting root mean 
square flux $S_{\rm rms}$ can be calculated in a straightforward manner 
\citep[cf.][]{intro.radio.astro};
\begin{equation}
\label{eq:sensitivity}
{S_{\rm rms} = \frac{2\,k_B T_{\rm sys}}{A_{\rm eff} 
    \sqrt{2\,\Delta \nu_{\rm rec} \tau}},}
\end{equation}
\noindent where $A_{\rm eff}$ is the total (effective) collecting area of the 
telescope, $T_{\rm sys}$ is the system temperature, $\Delta \nu_{\rm rec}$ is the
bandwidth used in the receiver, $\tau$ is the integration time and $k_B$ is Boltzmann's constant. 
The effective area $A_{\rm eff}$ and system temperature $T_{\rm sys}$ are the
key parameters. The SKA will have an effective area\footnote{Despite its
  name, it is unlikely that the SKA will have an area of 1 $\rm km^2$; 
  instead it is likely to be $\sim 0.5 \rm km^2$, which ensures greater survey 
  speed at the expense of sensitivity.} of order $A_{\rm eff}$=1 $\rm km^2$ and 
its pathfinders will have effective areas of a percentage of this; for 
pathfinders such as ASKAP, MeerKAT and APERTIF, this percentage will be 
$\lesssim 1\%$. A conservative estimate of the system temperature would be 
$T_{\rm sys}=50 K$. We can rewrite Eq.~(\ref{eq:sensitivity}) as

\begin{equation}
\label{eq:sensitivity_practical}
\frac{S_{\rm rms}}{1.626 \mu \rm Jy} = \left(\frac{A_{\rm eff}}{\rm km^2}\right)^{-1} 
\left(\frac{T_{\rm sys}}{50 \rm K}\right)
\left(\frac{\Delta \nu_{\rm rec}}{\rm MHz}\right)^{-1/2}
	\left(\frac{\tau}{\rm hr}\right)^{-1/2} ,
\end{equation}
\noindent where $1 {\rm Jy} = 10^{-26} {\rm Wm^{-2} \rm Hz^{-1}}$. 
The limiting flux sensitivity of the telescope $S_{\rm lim}$ is then 
$S_{\rm lim} = n_{\sigma}\, S_{\rm rms}$, where $n_{\sigma}$ defines the 
threshold for a galaxy to be reliably detected. Once we have fixed the 
integration time $\tau$ and the bandwidth $\Delta \nu_{\rm rec}$, we have the 
limiting sensitivity of our radio telescope. 

It is worth remarking on the relationship between the frequency bandwidth 
$\Delta \nu_{\rm rec}$ in Eq.~\ref{eq:sensitivity} and 
Eq.~\ref{eq:sensitivity_practical}, which is particular to the
radio telescope, and the frequency width $\Delta \nu$ in 
Eq.~\ref{eq:source_flux_mass} and Eq.~\ref{eq:source_flux_mass2}, 
which is set by the velocity width of the HI line $\Delta\,V_{\rm los}$. 
If we are to maximise the signal-to-noise ratio for detecting the
galaxy in a survey, then it is important that these bandwidths are matched. 
The overall telescope frequency bandwidth at a given frequency will be broad 
-- typically $\gtrsim 100$ MHz -- and much greater than the velocity width 
of an individual galaxy (e.g. $\sim\!1$ MHz corresponds to a 
$\sim\!200\,{\rm km}\,{\rm s}^{-1}$ 
galaxy), but the overall bandwidth consists of $\sim\!10^3$ to $10^4$ frequency 
channels that are much narrower than the expected frequency width of a galaxy.
Therefore a single telescope pointing will produce a huge data cube 
centred on a frequency $\nu$ with an overall bandwidth 
that consists of thousands of narrower frequency channels
$\delta\nu$. These channels will then be re-binned to produce data
cubes with different frequency resolutions $\Delta \nu_{\rm rec}$, and
one of these re-binnings will have $\Delta \nu_{\rm rec} \simeq \Delta
\nu$, which will be optimal for detecting an individual galaxy of a given
velocity width with a sufficiently high signal-to-noise.

\subsubsection{Angular Resolution}
\label{ssec:angres}

The angular resolution of the radio telescope becomes important when
the HI source is extended rather than a point source. For a single
dish telescope the angular resolution is $\sim \lambda/D$, where
$\lambda$ is the wavelength of the radiation and $D$ is the diameter
of the dish. Sources with angular sizes $\theta$ smaller than this are
indistinguishable from point sources. For radio interferometers it is
the lengths of the baselines between pairs of dishes $B$ rather than the
diameters of individual dishes that dictate the angular resolution. If
the longest baseline is $B_{\rm max}$ and the shortest is $B_{\rm
min}$, then the interferometer will resolve angular scales roughly
from $\Theta_{\rm min} = \lambda/B_{\rm max}$ to $\Theta_{\rm max} =
\lambda/B_{\rm min}$. Interferometers can therefore provide higher
angular resolution than a single dish, which is desirable because it
allows for HI sources to be mapped in greater detail. However, sources
more extended than $\sim\!\Theta_{\rm max}$ get resolved out, and have
their fluxes underestimated. The fraction of a galaxy's flux that is
resolved out will depend on, for example, the precise distribution of
interferometer baselines and what one assumes for the surface
brightness profile of the galaxy \citep[see, for example, the discussion
in][]{abdalla.etal.2009}. In the case of the SKA, the shortest
baseline is expected to be 20m, which corresponds to an angular resolution of
$\Theta_{\rm max} \sim$ 2100'' at $\lambda$=21 cm, while the maximum
baseline will be $\gtrsim 3000$km, which corresponds to a resolution
$\Theta_{\rm min} \sim$0.1''. As we will see in
\S\ref{ssec:observables}, the predicted HI sizes of galaxies in
cosmological surveys are typically of order arcsecs or smaller, so
there should not be any problem in practice with galaxies being
resolved out. 

\subsection{Predictions for Observables}
\label{ssec:observables}

\begin{figure*}
\centering
\includegraphics[width=8.0cm]{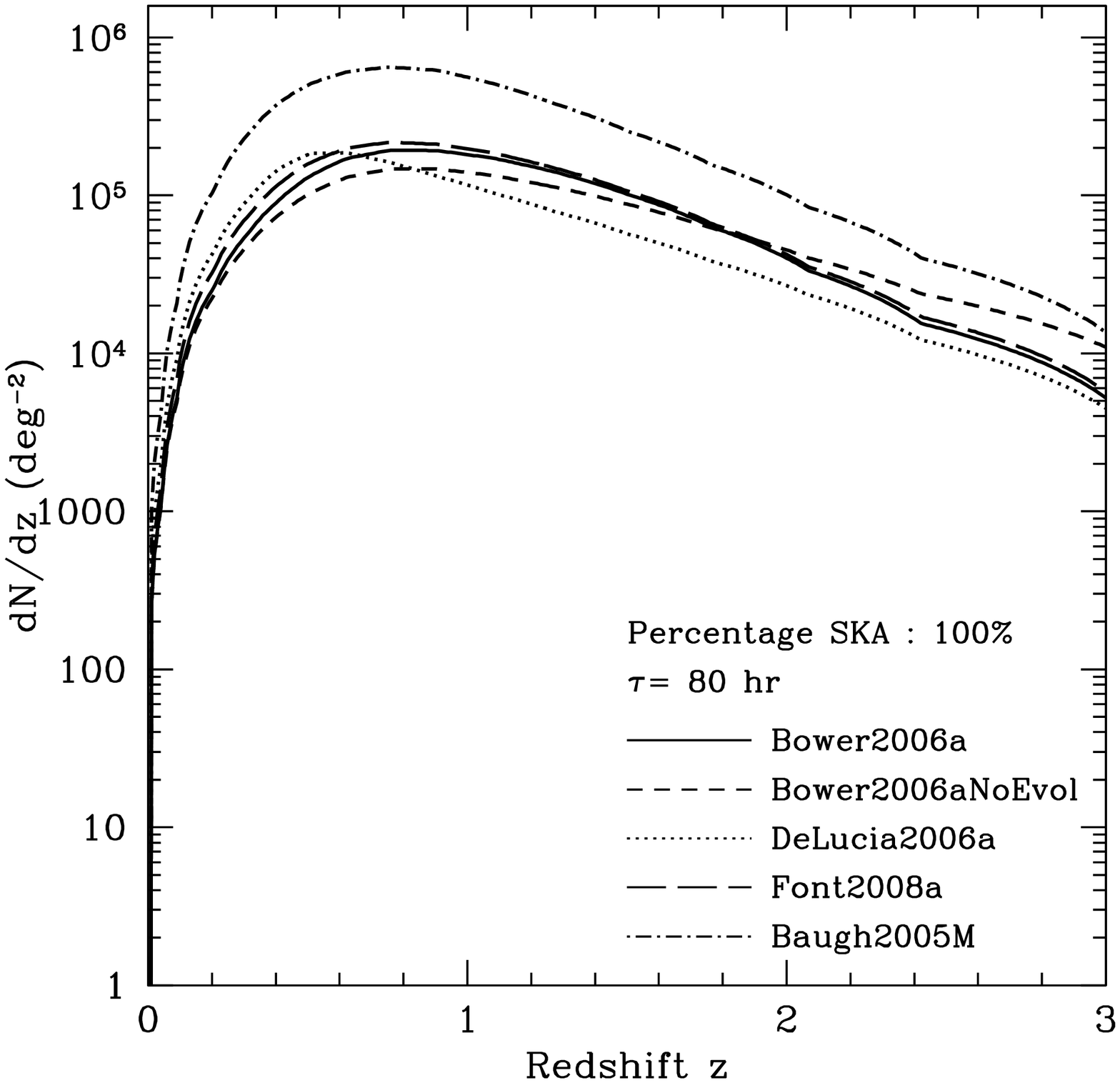}
\includegraphics[width=8.0cm]{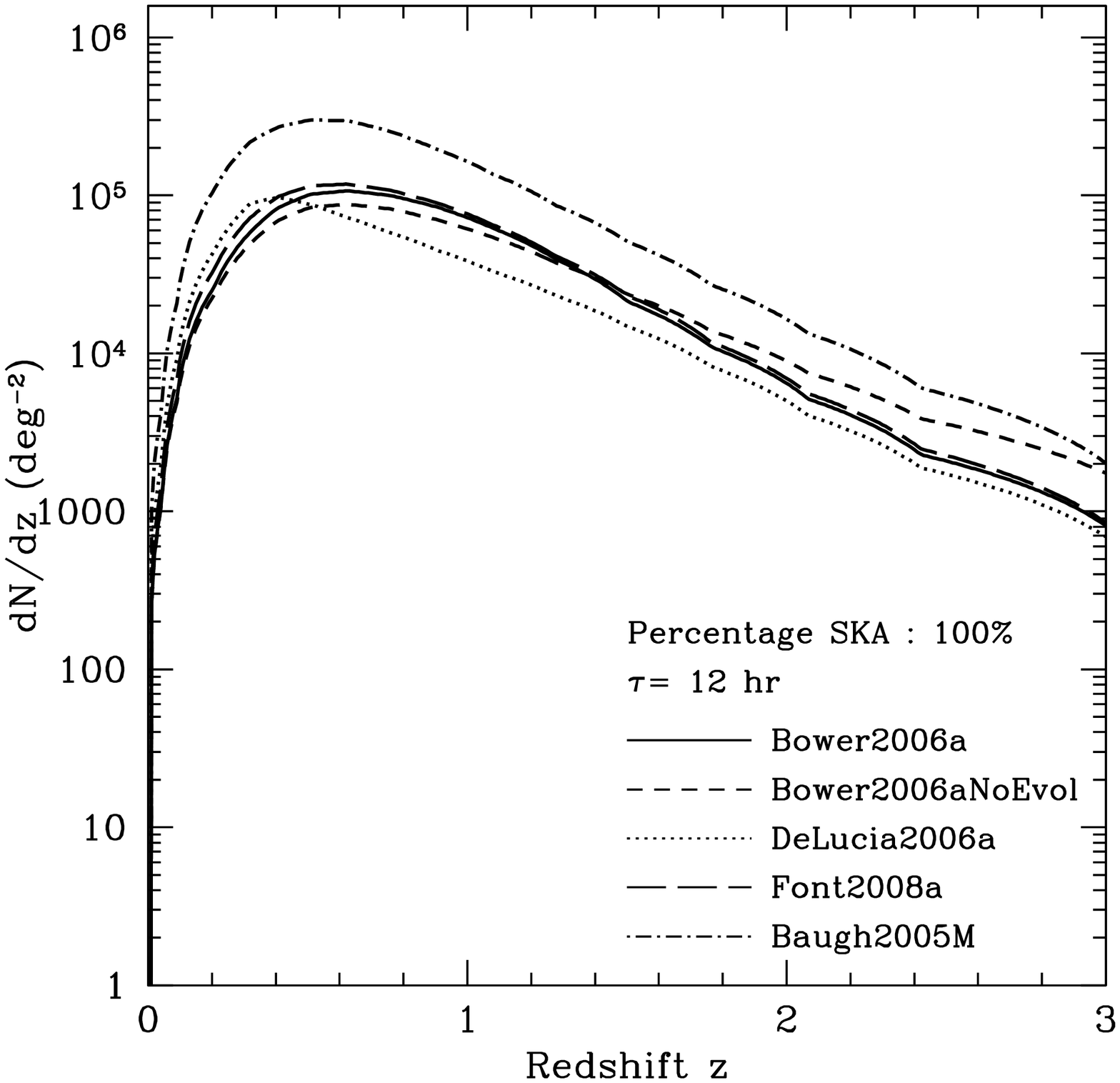}\\
\centering
\includegraphics[width=8.0cm]{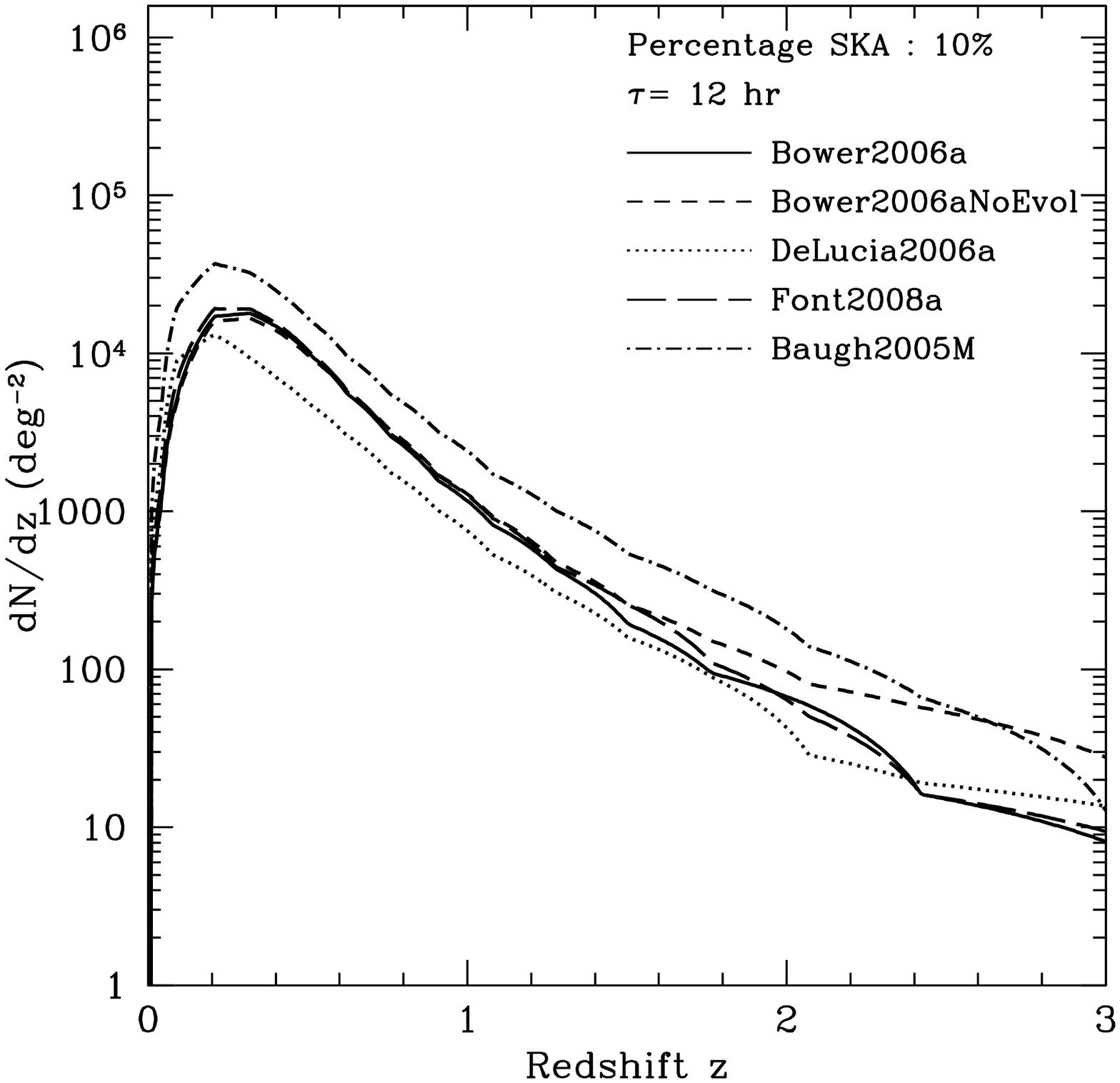}
\includegraphics[width=8.0cm]{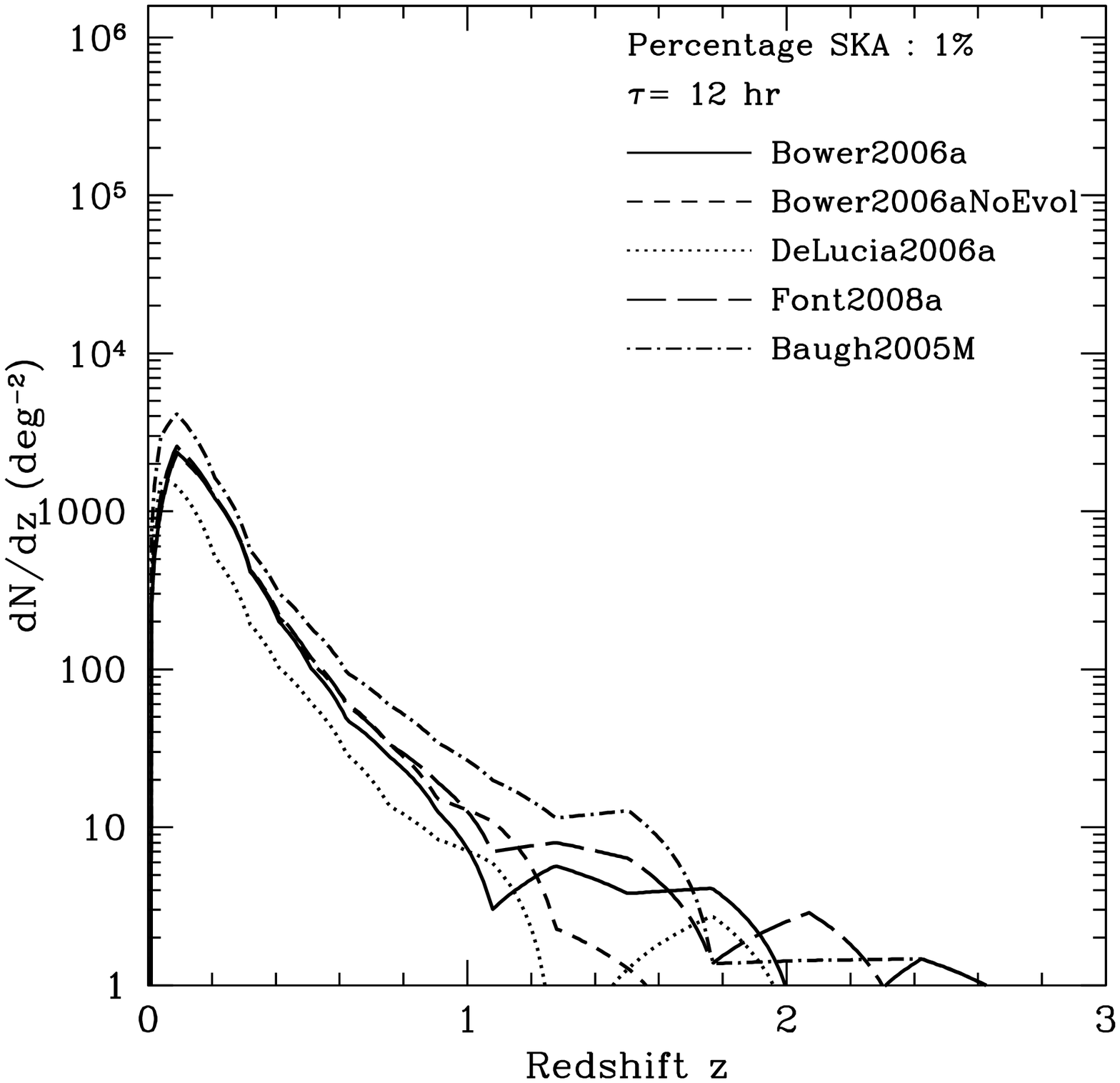}\\

\caption{Number counts of galaxies per square degree per unit redshift 
   for a telescope with 100\% (top left and right), 10\% (bottom left) and 
   1\% (bottom right) of the effective area of a fiducial SKA ($A_{\rm eff}$=1 
   $km^2$), for a deep peak flux limited hemispheric HI survey lasting 1 year. 
   We consider two integration times; $\tau$=12 hours, appropriate for a field 
   of view of 30 square degrees, which is typical of pathfinders such as ASKAP 
   and MeerKAT, and $\tau$=80 hours, appropriate for a field of view of 200 
   square degrees, which is the maximum field of view anticipated for the full
   SKA. Only sources that satisfy $S_{\rm obs} \geq n_{\sigma} S_{\rm rms}$ with 
   $n_{\sigma}$=10 are included. Note that we do not include galaxies with cold 
   gas masses below the resolution limit of $M_{\rm cold}= 10^{8.5} h^{-1} \, 
   \rm M_{\odot}$. In all of the panels we show how the counts change in 
   Bower2006a if we assume a ``No Evolution'' case in which the mass function 
   predicted for $z=0$ applies at all redshifts.}
\label{fig:dndz_vs_z}
\end{figure*}

First, we examine the predictions of the four models for the number counts 
${\rm d}N/{\rm d}z$ of HI galaxies per square degree of sky as a function of 
redshift. By number counts, we mean the number of HI sources ${\rm d}N$ that 
can be detected in a redshift interval ${\rm d}z$ centred on a redshift $z$,
\begin{equation}
\frac{{\rm d}N}{{\rm d}z} = \frac{{\rm d}V}{{\rm d}z} \int_0^\infty 
\frac{{\rm d}n}{{\rm d}M} f(M) \, {\rm d}M.
\label{eq:dndz}
\end{equation}
\noindent where ${\rm d}V/{\rm d}z$ is the cosmological volume element at 
redshift $z$, ${\rm d}n/{\rm d}M$ is the HI mass function at $z$ and $f(M)$ 
represents the fraction of galaxies with HI mass $M$ that can be detected by 
the radio telescope. For simplicity we assume that $f(M)$ depends only on 
limiting sensitivity, which depends on $M_{\rm HI}$. The angular resolution 
of the telescope also plays a role but its influence on $f(M)$ requires 
further assumptions to be made about, for example, the distribution of 
baselines, the clumpiness of HI within galaxy, its surface density profile, 
etc... and so we ignore this dependence.

In estimating predicted number counts, we assume a peak flux 
limited survey lasting 1 year on a radio telescope with an effective area 
$A_{\rm eff}$ of 1\%, 10\% and 100\% of the fiducial SKA\footnote{Assuming 
  that the SKA has an effective area of 1 $\rm km^2$, although as noted 
  already, the final SKA is likely to have an effective area smaller than 
  this.}. We make the simplifying assumption that the 
field of view is fixed with redshift and consider two cases -- 200 square
degrees, which could be achieved on the SKA \citep[cf.][]{taylor.2008} and 
30 square degrees, which is expected on ASKAP 
\citep[cf.][]{askap.science.2008}. Assuming that the survey covers a complete
hemisphere (i.e. $2\pi$ sr.), this gives effective integration times on a
patch of sky of $\tau=80$ and $\tau=12$ hours respectively. The measured flux 
$S_{\rm obs}$ from a galaxy is estimated using Eq.~\ref{eq:source_flux_mass2}. 
The velocity width $\Delta V_{\rm los}$ is taken to be $2 V_{\rm c,half} \sin i$,
where $V_{\rm c,half}$ is the circular velocity at the half-mass
radius of the galaxy. Galaxies are given random inclinations $i$,
drawn from a uniform distribution in $\cos i$. The measured flux is
compared to the limiting flux $S_{\rm rms}$
(Eq.~\ref{eq:sensitivity}) on a galaxy-by-galaxy basis (assuming
$\Delta\nu_{\rm rec}=\Delta\nu$ and using Eq.~\ref{eq:deltanu} to
estimate $\Delta\nu$) to estimate the signal-to-noise ratio. Our criterion
for detection is $S_{\rm obs}/S_{\rm rms} \geq n_{\sigma}=10$.

Fig.~\ref{fig:dndz_vs_z} shows how the number counts of HI galaxies varies
with redshift for surveys with $A_{\rm eff}$ of 100\% (top left 
panel for an integration of 80 hours, top right panel for an 
integration time of 12 hours), 10\% (bottom left panel) and 
1\% (bottom right panel) the effective area of the fiducial SKA ($A_{\rm eff}$=
$1 \rm km^2$). $A_{\rm eff}$ is crucial in determining how many 
galaxies can be detected and the range of redshifts that can be probed. 
We find the number counts peak at $\sim 4\times 10^3/4\times 10^4/
3\times 10^5$ galaxies per square degree at $z\!\sim$ 0.1/0.2/0.5 for a
year long HI hemispheric survey on a 1\%/10\%/100\% SKA with a 30 square 
degree field of view, corresponding to an integration time of 12 hours. On 
a full SKA with a 200 square degree field of view (equivalent to an 
integration time
of 80 hours) the number counts peak at $5\times 10^5$ galaxies per square 
degree at $z\!\sim$ 0.6.

A couple of interesting trends are immediately apparent in this figure. 
The first is that DeLucia2006a, Bower2006a and Font2008a, which all 
incorporate a form of AGN feedback, all predict broadly similar number counts 
out to 
$z \sim 3$. There are differences in the details that 
reflect differences between the models that can be readily inferred from 
the mass functions shown in Fig.~\ref{fig:mcold}. For example, 
DeLucia2006a predicts enhanced number counts at lower redshifts and
depressed number counts at intermediate to high redshifts with respect to 
Bower2006a and Font2008a. Here 
lower, intermediate and higher are defined relative to the redshift at which 
the number counts peak -- approximately $z \sim $0.5, 1.0 and 1.5 for 1\%, 
10\% and 100\% the effective area of the SKA respectively. The second is 
that Baugh2005M consistently predicts many more gas-rich galaxies than the 
other three models. There are several reasons for this: (1) Baugh2005M 
incorporates galactic super-winds rather than AGN feedback, which affects 
the cooling rate in massive haloes; (2) it uses weaker supernovae feedback 
than the other models; and (3) the star formation timescale in galactic discs 
does not scale with the disc dynamical time in Baugh2005M, whereas it does in 
the other models. \\

So far we have neglected the important issue of completeness of the
number counts. As explained in \S\ref{sec:results}, the finite mass 
resolution of the Millennium simulation implies that there is a minimum cold
gas mass -- and therefore a minimum HI mass -- that can be reliably resolved 
in the Millennium galaxy formation models. For our assumed cold gas mass 
limit of $10^{8.5} \rm M_{\odot}$, this implies a HI mass limit of 
$\sim 10^{8.2}\rm M_{\odot}$ (assuming a fixed $\rm H_2/HI$ approach to 
converting from cold gas mass to HI mass). The sensitivity of a survey may 
be such that galaxies with HI masses below this lower mass limit can be 
detected, and so we expect that the number counts of HI sources will be 
underestimated below some redshift $z_{\rm inc}$. This is because the cold 
gas mass function has not converged at low masses and so the population of 
sources is incomplete. As the sensitivity of a survey increases, so too does 
$z_{\rm inc}$ because the survey probes the HI mass function down to lower 
masses and the effect of this incompleteness will be in evidence at higher 
redshifts. 

We have assessed the issue of completeness for one of the Durham models 
(Bower2006a) using three sets of merger trees of increasing resolution 
(1, 2 and 4 times the Millennium galaxy formation model resolution; 
MC1, MC2 and MC3 respectively) generated using the Monte Carlo prescription 
described in \S\ref{sec:results}. We considered surveys with effective 
areas of 1\%/10\%/100\% of the SKA and integration times of 12 and 80 hours. 
As we would expect, the number counts predicted using the $N$-body merger 
trees and the MC1 merger trees are consistent with one another; reassuringly,
the same holds true for the MC2 and MC3 merger trees, for all of the survey 
sensitivities that we considered. This suggests that the number counts we
obtain for Bower2006a with the $N$-body trees are converged and the peak 
redshifts and number counts we find in Fig.~\ref{fig:dndz_vs_z} are robust. 
We have also estimated the redshifts at which the limiting survey sensitivity
(and consequently minimum HI mass detectable; see 
Eq.~\ref{eq:sensitivity}) is comparable to the limiting
HI mass and we find that $z_{\rm inc} \lesssim 0.04$ for a 1\%/10\% SKA, and 
$z_{\rm inc} \simeq 0.08$ for a 100\% SKA. These estimates confirm that 
incompleteness effects are unlikely to affect the shape and amplitude of the 
peak of the number counts shown in Fig.~\ref{fig:dndz_vs_z} for 
Bower2006a\footnote{It is worth remarking that the precise 
  value of $z_{\rm inc}$ will vary from model to model because each model 
  predicts slightly different shapes for the HI mass function with 
  decreasing mass, and so we should repeat this exercise for each model 
  in principle. However, the consistency of Bower2006a with the other 
  models suggests that conclusions about completeness based on this model 
  are sufficiently general to be applied to the other models.}\\

\begin{figure}
\centering
\includegraphics[width=8.0cm]{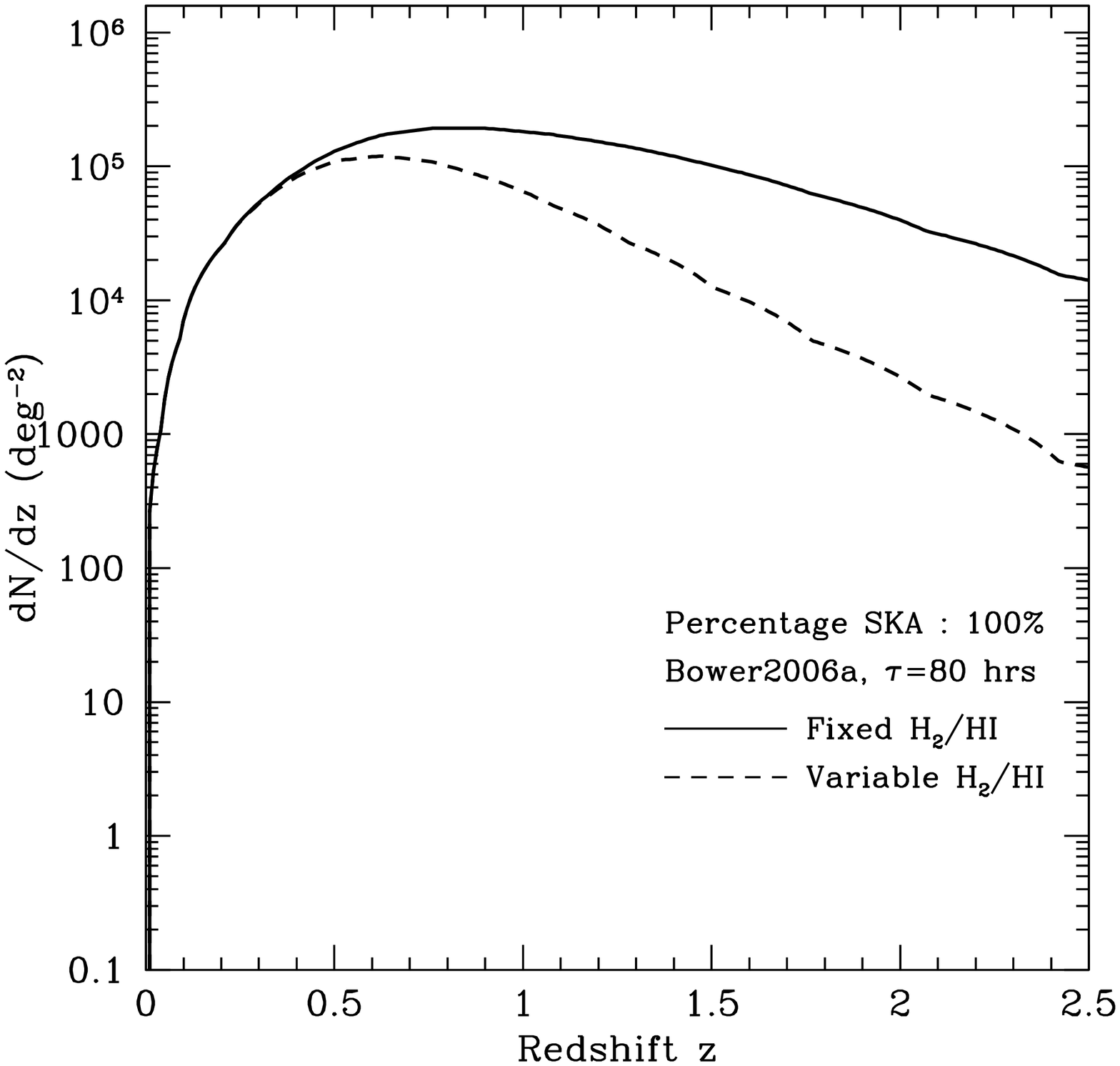}
\includegraphics[width=8.0cm]{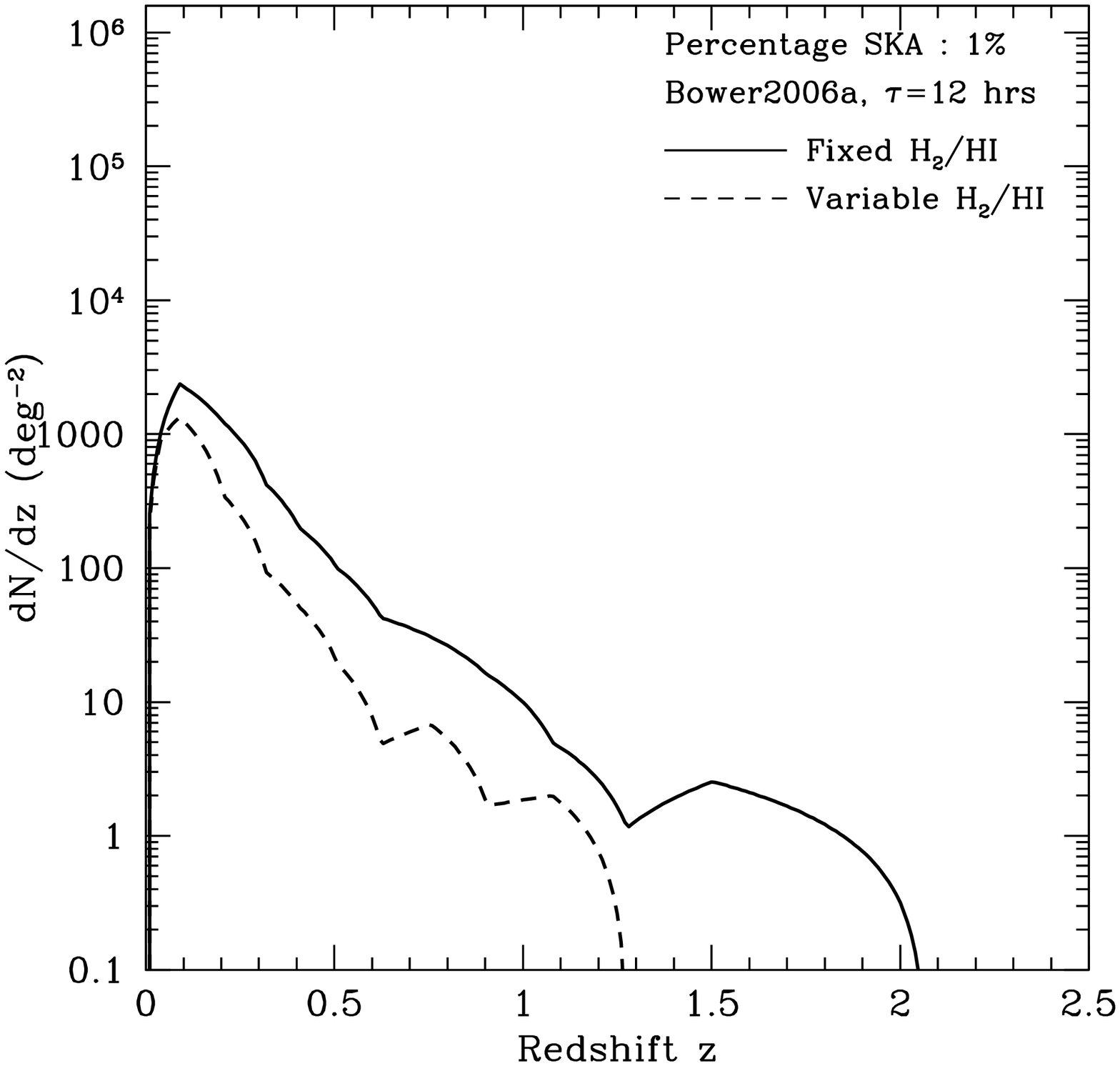}
\caption{Number counts of galaxies per square degree per unit 
  redshift for a telescope with 100\% (top) and 1\% 
  (bottom) of the effective area of a fiducial SKA (1 $\rm km^2$), 
  for a peak flux limited HI survey lasting 1 year, based on
  Bower2006a. The integration times are $\tau$=80 hrs and 12 hrs respectively.
  As before, only sources that satisfy
  $S_{\rm obs} \geq n_{\sigma} S_{\rm rms}$ with $n_{\sigma}$=10 are included
  and we ignore galaxies with cold gas masses below the 
  resolution limit. The solid and dashed curves correspond to the (fiducial) 
  fixed $\rm H_2/HI$ and variable $\rm H_2/HI$ approaches to 
  estimating the cold gas mass to HI mass conversion factor.}
\label{fig:dndz_variable}
\end{figure}

Fig.~\ref{fig:dndz_vs_z} is based on the fixed $\rm H_2/HI$ approach to 
estimating HI masses from cold gas masses, but it is interesting to ask how
the number counts would change if we used the variable $\rm H_2/HI$ approach
instead. In Fig.~\ref{fig:dndz_variable} we compare the number 
counts of sources in Bower2006a estimated using the fixed $\rm H_2/HI$ 
approach (solid curves) and the variable $\rm H_2/HI$ approach (dashed curves).
In the upper panel we show the case for a 100\% SKA with an integration time of
$\tau=80 \rm hrs$ and in the lower panel we show the result for a 1\% SKA with 
an integration time of $\tau=12 \rm hrs$. As discussed in 
\S\ref{ssec:convert_cg_to_hi}, we expect the number of sources to be 
systematically lower at higher redshifts if we adopt a variable rather than a 
fixed $\rm H_2/HI$ ratio and this is confirmed by Fig.~\ref{fig:dndz_variable}.
The peak number of sources is lower in the variable approach and the number 
counts decline more sharply with increasing redshift. The difference is a 
factor of 10 for the 100\% (1\%) of the SKA by $z$=3 (1). \\

In Fig.~\ref{fig:mcold} we noted that the cold gas mass function does not appear
to evolve strongly with redshift between $0 \leq z \lesssim 1$. It is therefore
interesting to ask how approximating the mass function at $z$ by the mass
function predicted at $z$=0 impacts on the number counts of sources. In each of
the panels in Fig.~\ref{fig:dndz_vs_z} the solid short dashed curves highlight 
the impact making such an approximation -- showing the behaviour of the 
Bower2006aNoEvol approximation, in which the HI mass function at a given $z$ is
replaced by the HI mass function predicted by Bower2006a at $z$=0. For the 10\%
and 100\% SKA this would appear to be a reasonable approximation over 
the redshift range $0 \lesssim z \lesssim 1.5$, differing by $\sim 10$\% at 
most. At $z \gtrsim 1.5$ the predicted number counts diverge, the degree of the
discrepancy depending on the sensitivity of the survey.\\

It's useful to compare our results with previous work, and so we 
note that \citet{abdalla.rawlings.2005} have predicted the redshift 
variation of ${\rm d}N/{\rm d}z$ for a full SKA using semi-empirical 
models for the HI mass function. This is interesting because we can compare 
predictions based on semi-analytical models with their predictions based on 
semi-empirical models. For a survey with an integration time of 4 hours on a 
full SKA, they expect to detect $\sim 8 \times 10^4$ sources at the peak 
${\rm d}N/{\rm d}z$; this peak occurs at $z\sim 0.6$. Based on a similar 
integration time, the semi-analytical models predict peak ${\rm d}N/{\rm
  d}z$'s of between $\sim 8 \times 10^4$ (Bower2006a,DeLucia2006a,Font2008)
and $\sim 2 \times 10^5$ (Baugh2005M). All of these models peak at 
$z\sim 0.5$. The semi-analytical models predict ${\rm d}N/{\rm d}z$'s that 
decline more gently with increasing $z$ than the semi-empirical models but 
this reflects in part differences in the model assumptions (e.g. the 
assumption that HI, baryons and dark matter follow similar mass functions) 
and in part the conversion from cold gas to HI mass that we must assume.\\

\begin{figure}
\includegraphics[width=8cm]{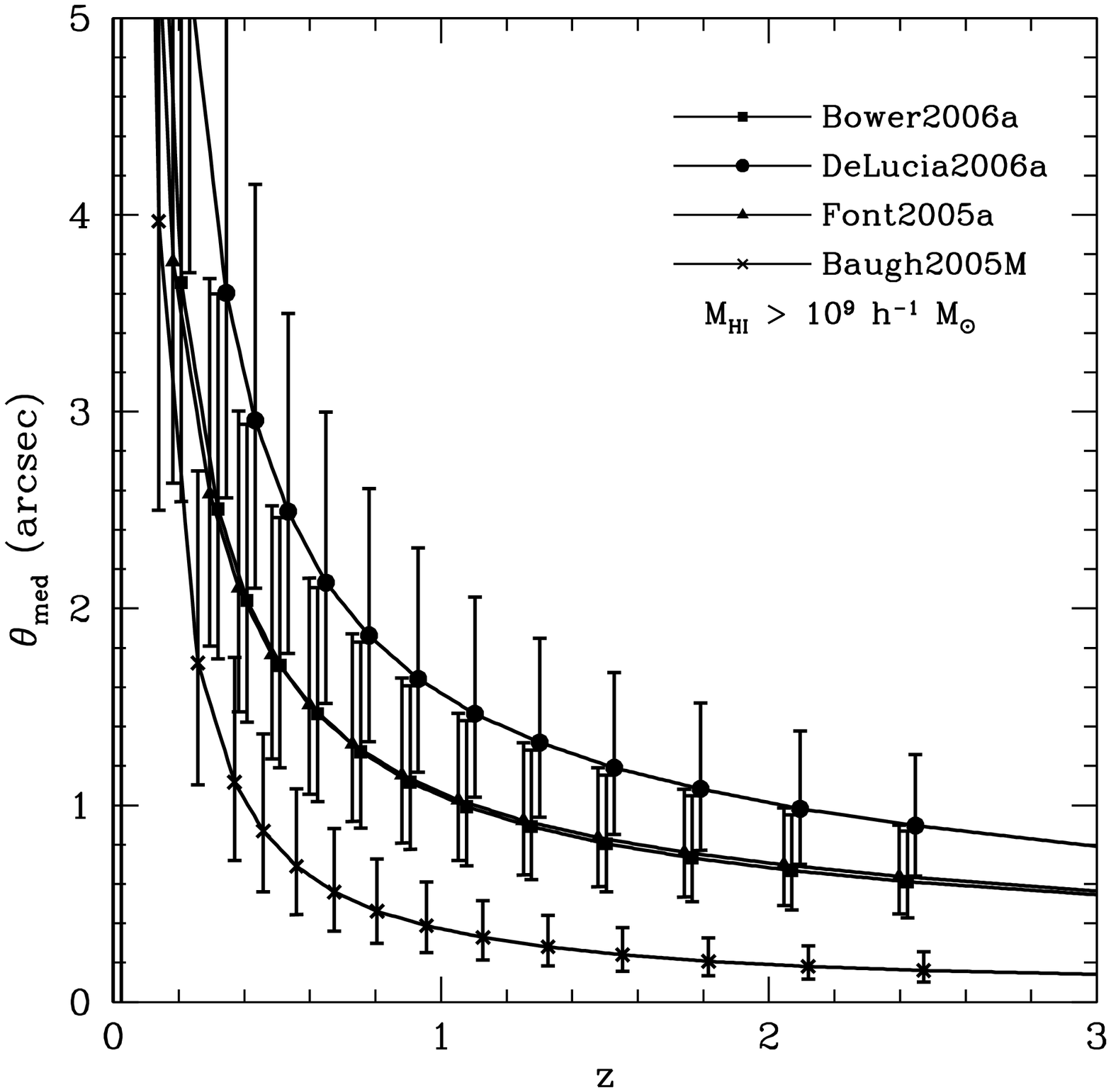}
\includegraphics[width=8cm]{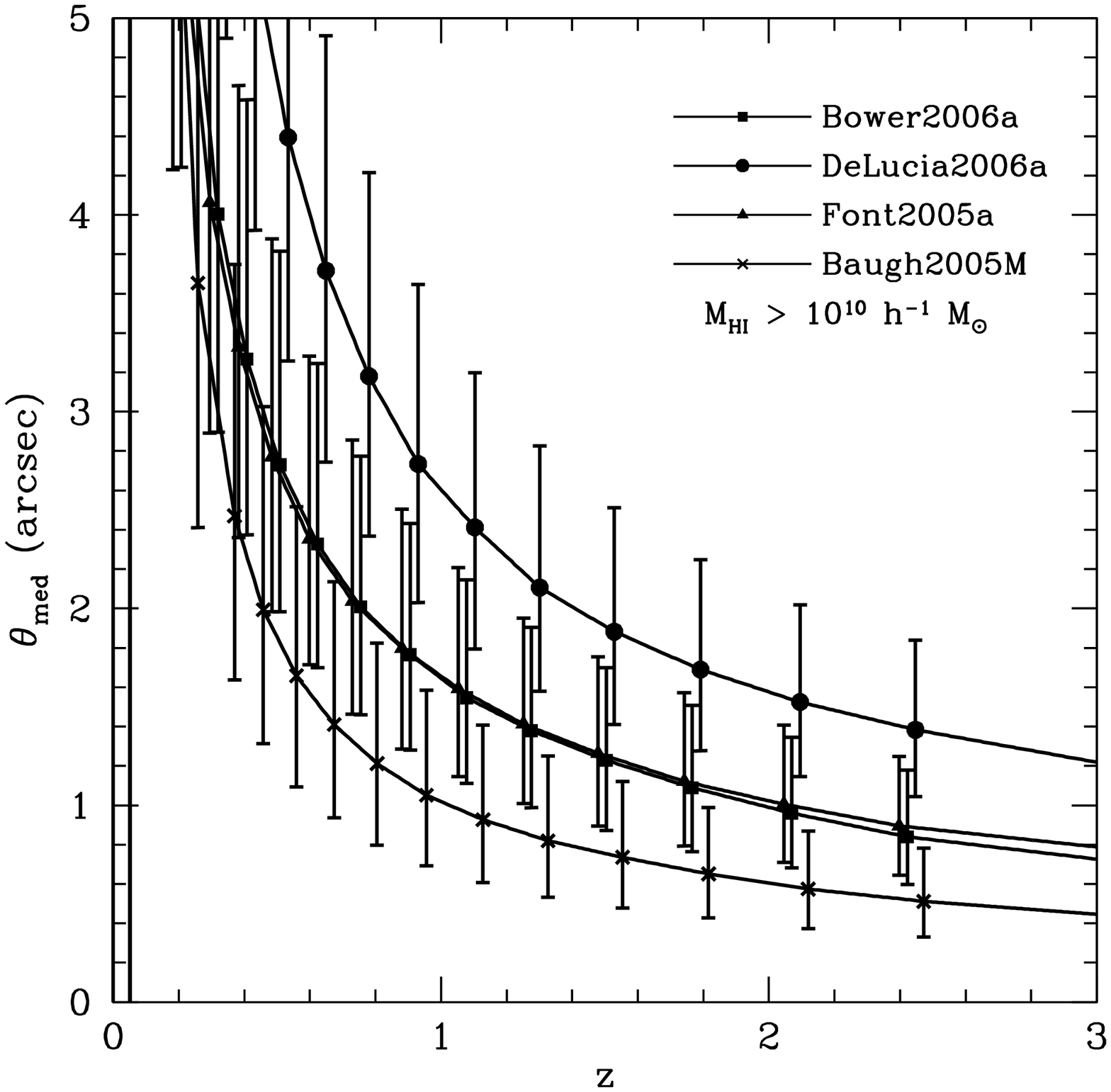}
\caption{The predicted redshift variation of the angular diameter of 
  galaxies with HI masses $M_{\rm HI}\gtrsim 10^9 h^{-1} \rm M_{\odot}$
  (upper panel) and $M_{\rm HI}\gtrsim 10^{10} h^{-1} \rm M_{\odot}$ (lower
  panel). Different symbols correspond to different models, as indicated by
  the legend.}
\label{fig:ang_diameter}
\end{figure}

\begin{figure}
\includegraphics[width=8cm]{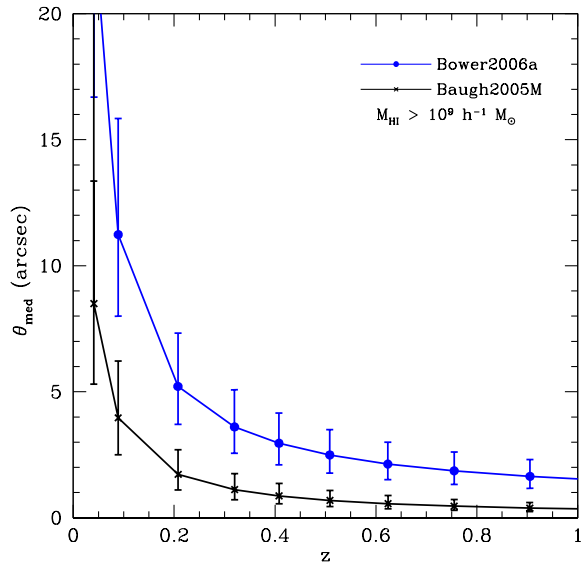}
\caption{The predicted redshift variation of the angular diameter of 
  galaxies with HI masses $M_{\rm HI}\gtrsim 10^9 h^{-1} \rm M_{\odot}$ at $z
  \lesssim 1$ in Baugh2005M and DeLucia2006a.}
\label{fig:ang_diameter_lowz}
\end{figure}

Second, we focus on the angular sizes of galaxies. 
Fig.~\ref{fig:ang_diameter} and Fig.~\ref{fig:ang_diameter_lowz} show 
how the angular size of HI galaxies varies with redshift. We compute the 
angular diameter as $\theta=2\,R_{\rm h}/D_{\rm ang}$ where $R_{\rm h}$ is the 
half-mass radius of the galaxy and $D_{\rm ang}(z)=(1+z)^{-1}\,D_{\rm co}(z)$
is the angular diameter distance of the galaxy with respect to the observer,
where, as before, $D_{\rm co}(z)$ is the radial comoving separation between 
source and observer. The points correspond to the median angular diameters 
while the upper and lower error-bars indicate the $25^{\rm th}$ and $75^{\rm
  th}$ percentiles of the angular diameter distributions at that redshift. 
Points are given horizontal offsets of 0.025 in redshift to aid clarity.
In Fig.~\ref{fig:ang_diameter} we plot the redshift dependence of the median 
angular diameters of galaxies with HI masses $M_{\rm HI} \geq 10^9 h^{-1} 
\rm M_{\odot}$ (upper panel) and $M_{\rm HI} \geq 10^{10} h^{-1} \rm
M_{\odot}$ (lower panel) varies with redshift out to $z \lesssim 3$. In
Fig.~\ref{fig:ang_diameter_lowz} we focus on the variation predicted by
Baugh2005 and DeLucia2006a for galaxies with $M_{\rm HI} \geq 10^9 h^{-1} 
\rm M_{\odot}$ over the redshift interval $0 \leq z \leq 1$. 

Knowledge of the expected redshift variation of angular diameter of an
extended HI source is useful because it allows one to estimate the number 
of sources that are likely to be resolved out. 
Fig.~\ref{fig:ang_diameter} and Fig.~\ref{fig:ang_diameter_lowz} distill 
the information presented in
Fig.~\ref{fig:rdisc}, where we showed how the half-mass radii of galaxies
varied with cold gas mass at a given redshift. The models indicate that
galaxies that have larger HI masses tend to have larger half-mass
radii, which corresponds to more extended angular diameters at a given
redshift. The angular diameter decreases sharply between $z$=0 and $z \simeq$
0.5, and more gently at $z\gtrsim 0.5$. The median angular size
varies between $5''$ and $10''$ at $z$=0.1, $1''$ and $3''$ at $z$=1 and 
$1''$ and $3''$ at $z$=3 for galaxies with HI masses in excess of 
$10^9 h^{-1} \rm M_{\odot}$; the upper and lower bounds correspond to the
predictions from DeLucia2006a and Baugh2005M. Therefore, to resolve a 
typical galaxy with an HI mass of $M_{\rm HI}\gtrsim 10^9 h^{-1} \rm
M_{\odot}$ at $z \sim$ 1 requires a maximum baseline of order $100$ km.

\section{Summary}
\label{sec:summary}

Neutral atomic hydrogen is the fundamental baryonic building block of galaxies
and understanding how its abundance varies over cosmic time will provide us
with important insights into galaxy formation. Few observational data exist 
for the abundance of neutral hydrogen at redshifts 
$z \gtrsim 0.05$, the extent of the HIPASS survey \citep{meyer.etal.2004}, but 
this will change with the advent of the Square Kilometre Array, which will see
first light by about 2020. The SKA will transform cosmology and galaxy formation
\citep[e.g.][]{blake.etal.2007,braun.2007}, allowing us to probe 
the cosmic HI distribution out to redshifts $z\!\sim\!3$. In the meantime,
HI surveys on SKA pathfinders such as ASKAP \citep{askap.science.2008},
MeerKAT \citep{meerkat.2007} and APERTIF \citep{apertif.2008} will provide 
us with important initial glimpses into the cosmic HI distribution out to 
redshifts $z\!\sim\!1$. Because we have such few observational data for the 
cosmic HI distribution beyond $z\!\sim\!0.05$, the coming decade promises to 
provide powerful tests of the predictions of theoretical galaxy formation 
models. It is therefore timely to ask what galaxy formation models tell us about
the abundance of neutral hydrogen in galaxies. 

In this paper we have investigated four of the currently favoured galaxy 
formation models -- those of \citet{baugh.etal.2005},
\citet{bower.etal.2006}, \citet{delucia.blaizot.2007} and 
\citet{font.etal.2008} -- and determined what
they predict for the mass function of cold gas in galaxies and how it evolves
with redshift. Each of the models use merger trees derived from the Millennium 
simulation \citep[cf.][]{springel.etal.2005} and so any differences between
the model predictions reflect intrinsic differences in the 
physics incorporated into the models themselves. Three of the models 
(Baugh2005M, Bower2006a and Font2008a) use the
Durham semi-analytical code {\small GALFORM} \citep[cf.][]{cole.etal.2000}
whereas the fourth (DeLucia2006a) uses the Munich semi-analytical
code. Arguably the most important difference between the models is in the
precise treatment of feedback; Bower2006a, DeLucia2006a and Font2008a all 
incorporate a form of AGN feedback whereas Baugh2005M does not, instead 
favouring galactic super-winds.

Interestingly we find that the model predictions are broadly
consistent with one another. Differences between the models reflect 
(1) the use of AGN heating to suppress gas cooling in
massive haloes (which is used in Bower2006a, Font2008a and
DeLucia2006a but not in Baugh2005M, which invokes supernovae-driven
super-winds); (2) the strength of supernovae feedback, which is weakest in
Baugh2005M; and (3) the scaling (or lack of) of the star formation
timescale in galactic discs with the disc dynamical time (scaling is assumed in
Bower2006a, DeLucia2006a and Font2008a whereas it is not in Baugh2005M).

We have focused on three particular aspects of the cold gas
properties of galaxies, namely (i) the mass function of cold gas in galaxies and
the relationship between (ii) a galaxy's cold gas mass and its half-mass radius
and (iii) its cold gas mass and rotation speed (i.e. circular velocity) at 
this radius.

\vspace{0.25cm}

\noindent \emph{Mass function of cold gas in galaxies}: The predictions of 
Font2008a and Bower2006a are generally very similar, with differences 
only apparent at small cold gas masses. This is unsurprising because 
Font2008a descends directly from Bower2006a, the
principal difference between the models being the improved treatment 
of gas stripping by the hot intra-cluster medium in Font2008a. At $z$=0 
we find that Bower2006 and Font2008 systematically over-predict the numbers 
of galaxies with HI masses in excess of $10^9 h^{-2} \rm M_{\odot}$ when 
compared to the observed mass function derived from HIPASS 
\citep[cf.][]{zwaan.etal.2005}, while Baugh2005M and DeLucia2006a provide
reasonable descriptions (i.e. in terms of shape and amplitude) of the observed
mass function. Interestingly we find that the cold gas mass function shows 
little evolution out to redshifts of $z \simeq 3$ in all four models.

\vspace{0.2cm}

\noindent \emph{Relationship between cold gas mass and half-mass radius,
  rotation speed}: At fixed cold gas mass, both Bower2006a and 
Font2008a predict half-mass radii and rotation speeds (circular velocities 
measured at these half-mass radii) that are in excellent agreement with each 
other, as we might expect. Half-mass radii are slightly but systematically 
larger (by $\sim 25\%$) in DeLucia2006a than in Bower2006a and Font2008a, 
but the three models predict similar rotation speeds (to within the width of 
the distribution). This level of agreement is remarkable, given the number of 
subtle (and some not so subtle) differences between the frameworks underpinning
DeLucia2006a and Bower2006a/Font2008a (described in \S\ref{sec:galform}). 
In contrast, Baugh2005M predicts half-mass radii that are systematically 
smaller ($\sim 60\%$) and rotation speeds that are systematically larger 
(by $\sim 20\%$) at fixed cold gas mass than in Bower2006a, DeLucia2006a and 
Font2008a. It is worth noting that Baugh2005M predicts a size-luminosity 
relation for late-type galaxies that is in very good agreement with SDSS data, 
whereas the agreement between Bower2006a (and by extension Font2008a) and the 
observational data is poor \citep[cf.][]{gonzalez.etal.2008}.
\vspace{0.25cm}

We took the predicted mass functions of cold gas in galaxies and used
them to derive number counts of HI galaxies for future all-sky HI
surveys. Rather than adopting a specific design, we considered surveys carried
out on radio telescopes with effective collecting areas $A_{\rm eff}$ that are 
percentages of a fiducial Square Kilometre Array, with an effective 
collecting area of $1 \rm km^2$. We focused on surveys with $A_{\rm eff}$ of 
1\%, 10\% and 100\% of the SKA and assumed that these surveys lasted for 1 
year, with integration times of between 12 and 80 hours within individual 
fields of view. As we pointed out in \S\ref{ssec:sensitivity}, $A_{\rm eff}$ 
plays a crucial role in determining the sensitivity $S_{\rm rms}$ of a radio 
telescope, which in turn dictates how many HI galaxies are likely to be 
detectable by the survey. SKA pathfinders such as ASKAP, MeerKAT and APERTIF 
will have effective areas of order $\sim$ 1\%. 

We examined two possible approaches to converting cold
gas masses to HI masses. The first simply assumed that the ratio of 
molecular-to-atomic hydrogen ($\rm H_2/HI$) is fixed for all galaxies at 
all redshifts \citep[cf. ][]{baugh.etal.2004}. The second assumed 
that the $\rm H_2/HI$ ratio is variable, depending on individual galaxy
properties according to the model of \citet{obreschkow2009b}, which in
turn is based on an empirical relation between the ratio of the surface 
densities of $\rm H_2$ to $\rm HI$ and the gas pressure found for local 
galaxies by \citet{blitz.2006} and \citet{leroy.etal.2008}. This is an important
consideration because how one converts from cold gas mass to HI mass will 
determine the 21-cm luminosity of a galaxy and therefore its detectability in 
an HI survey of a given sensitivity. We computed the observed flux 
$S_{\rm obs}$ for each galaxy using both its HI mass and its 
circular velocity at the half-mass radius to define its velocity width and 
required that $S_{\rm obs} \geq 10 S_{\rm rms}$ for the galaxy to be
detected. 

As for the cold gas mass functions, we find that the models that include
a form of AGN feedback predict broadly similar number counts;
Baugh2005M predicts many more gas rich galaxies, as many as a factor
of $\sim 2$ --$3$ more at the redshift at which the number counts
peak. The choice of cold gas to HI mass conversion factor is 
also very important, especially at higher redshifts; adopting a variable 
$\rm H_2/HI$ ratio predicts that galaxies should be predominantly molecular 
rather than atomic hydrogen at high redshifts, and this has a profound impact 
on the number of HI sources one predicts (see Fig.~\ref{fig:dndz_variable}). 
Clearly more work is needed to put this on a more secure theoretical footing. 
Interestingly we find that approximating the HI mass function at 
$z \lesssim 2$ by the $z$=0 HI mass function has little impact on the number 
counts one might expect to measure. 

In addition, we estimated the dependence of the median angular diameter of HI
galaxies on redshift, for galaxies with HI masses 
$M_{\rm HI} \geq 10^9 h^{-1} \rm M_{\odot}$ and $M_{\rm HI} \geq
10^{10} h^{-1} \rm M_{\odot}$. This is useful to know because it allows one to 
estimate the fraction of the flux that is likely to be lost because it has
been resolved out. The models indicate that
galaxies with larger HI masses tend to have larger half-mass
radii and therefore more extended angular diameters at a given
redshift. We found that the angular diameter decreases sharply between 
$0 \lesssim z \lesssim 0.5$ and more gently at $z \gtrsim 0.5$. The median 
angular size varies between $5''$ and $10''$ at $z$=0.1, $0.5''$ and $3''$ 
at $z$=1 and $0.2''$ and $1''$ at $z$=3 for galaxies with HI masses in excess 
of $10^9 h^{-1} \rm M_{\odot}$, where the lower and upper limits correspond to
the predictions of Baugh2005M and DeLucia2006a.\\

We have concentrated on the most straightforward measurement one can make in
future HI surveys, namely the number counts of galaxies. However, we have 
considered only global counts -- we have not considered how the HI mass
functions and number counts might depend on local environment. Certainly there 
is good reason to expect that environment should play a role in shaping the HI
mass function of galaxies. For example, we might expect that the amount of
HI in a galaxy will be reduced by ram pressure stripping as it falls 
through a dense intra-cluster medium; this would become apparent as a quenching 
of the star formation \citep[e.g.][]{balogh.etal.2000,quilis.etal.2000}, but it
should also be evident in an environmental dependence of a galaxy's HI 
properties. Indeed, there is some observational evidence to suggest that 
the HI mass function does depend on environment 
\citep[e.g.][]{zwaan.etal.2005,springob.etal.2005,kilborn.etal.2009}; for
example, \citet{kilborn.etal.2009} find evidence that the slope of the low-mass
end of the HI mass function in galaxy groups decreases with decreasing HI mass,
in contrast to the global HI mass function found in HIPASS by 
\citet{zwaan.etal.2005}. 

These results are interesting because they suggest that environment plays an 
important role in determining the HI properties of galaxies. In forthcoming 
papers we will explore precisely what galaxy formation models predict for the 
clustering of cold gas (Kim et al., in preparation) and we will explore 
precisely what role environment plays in shaping a galaxy's cold gas -- and
consequently HI -- properties.

\section*{Acknowledgements}

We thank the referee Danail Obreschkow for their very helpful comments.
CP thanks Chris Blake, John Helly and Lister Staveley-Smith for instructive 
discussions at various stages during the writing of this paper. This work was 
supported by separate STFC rolling grants at Leicester and Durham. CP 
acknowledges the
support of the Australian Research Council funded ``Commonwealth Cosmology 
Initiative'', DP Grant No. 0665574 during the initial stages of this work.
We acknowledge the efforts of Andrew Benson, Richard Bower, Shaun
Cole, Carlos Frenk, John Helly and Rowena Malbon in developing the GALFORM
code used in the Baugh2005M model. This project was made possible by the
availability of models on the Millennium archive set up the Virgo Consortium
with the support of the German Astrophysical Virtual Observatory.

\bibliographystyle{mn2e}

\begin{thebibliography}{}

\bibitem[\protect\citeauthoryear{{Abdalla}, {Blake} \& {Rawlings}}{{Abdalla}
    et~al.}{2009}]{abdalla.etal.2009}
  {Abdalla} F.~B.,  {Blake} C. \& {Rawlings} S.,  2009, preprint (arXiv:astro-ph/0905.4311)
  
\bibitem[\protect\citeauthoryear{{Abdalla} \& {Rawlings}}{{Abdalla} \& 
    Rawlings}{2005}]{abdalla.rawlings.2005}
  {Abdalla} F.~B. \& {Rawlings} S.,  2005, \mnras, 360, 27 
  
\bibitem[\protect\citeauthoryear{{Almeida}, {Baugh} \& {Lacey}}{{Almeida}
    et~al.}{2007}]{almeida.etal.2007}
  {Almeida} C.,  {Baugh} C.~M. \& {Lacey} C.~G.,  2007, \mnras, 376, 1711
  
\bibitem[\protect\citeauthoryear{{Almeida}, {Baugh}, {Wake}, {Lacey}, {Benson},
    {Bower} \& {Pimbblet}}{{Almeida} et~al.}{2008}]{almeida.etal.2008}
  {Almeida} C.,  {Baugh} C.~M.,  {Wake} D.~A.,  {Lacey} C.~G.,  {Benson} A.~J.,
  {Bower} R.~G. \& {Pimbblet} K.,  2008, \mnras, 386, 2145
  
\bibitem[\protect\citeauthoryear{{Balogh}, {Navarro} \& {Morris}}{{Balogh}
    et~al.}{2000}]{balogh.etal.2000}
  {Balogh} M.~L.,  {Navarro} J.~F. \&  {Morris} S.~L.,  2000, \apj, 540, 113
  
\bibitem[\protect\citeauthoryear{{Baugh}}{{Baugh}}{2006}]{baugh.2006}
  {Baugh} C.~M.,  2006, Reports of Progress in Physics, 69, 3101
  
\bibitem[\protect\citeauthoryear{{Baugh}, {Lacey}, {Frenk}, {Benson}, {Cole},
    {Granato}, {Silva} \& {Bressan}}{{Baugh} et~al.}{2004}]{baugh.etal.2004}
  {Baugh} C.~M.,  {Lacey} C.~G.,  {Frenk} C.~S.,  {Benson} A.~J.,  {Cole} S.,
  {Granato} G.~L.,  {Silva} L. \&  {Bressan} A.,  2004, New Astronomy Review,
  48, 1239
  
\bibitem[\protect\citeauthoryear{{Baugh}, {Lacey}, {Frenk}, {Granato}, {Silva},
    {Bressan}, {Benson} \& {Cole}}{{Baugh} et~al.}{2005}]{baugh.etal.2005}
  {Baugh} C.~M.,  {Lacey} C.~G.,  {Frenk} C.~S.,  {Granato} G.~L.,  {Silva} L.,
  {Bressan} A.,  {Benson} A.~J. \& {Cole} S.,  2005, \mnras, 356, 1191
  
\bibitem[\protect\citeauthoryear{{Benson}, {Bower}, {Frenk}, {Lacey}, {Baugh}
    \& {Cole}}{{Benson} et~al.}{2003}]{benson.etal.2003.b}
  {Benson} A.~J.,  {Bower} R.~G.,  {Frenk} C.~S.,  {Lacey} C.~G.,  {Baugh}
  C.~M. \& {Cole} S.,  2003, \apj, 599, 38
  
\bibitem[\protect\citeauthoryear{{Benson}, {Frenk}, {Baugh}, {Cole} \&
    {Lacey}}{{Benson} et~al.}{2001}]{benson.etal.2001.a}
  {Benson} A.~J.,  {Frenk} C.~S.,  {Baugh} C.~M.,  {Cole} S. \& {Lacey} C.~G.,
  2001, \mnras, 327, 1041
  
\bibitem[\protect\citeauthoryear{{Blake}, {Abdalla}, {Bridle} \&
    {Rawlings}}{{Blake} et~al.}{2004}]{blake.etal.2007}
  {Blake} C.~A.,  {Abdalla} F.~B.,  {Bridle} S.~L. \& {Rawlings} S.,  2004, New
  Astronomy Review, 48, 1063
  
\bibitem[\protect\citeauthoryear{{Blitz} \& {Rosolowsky}}{{Blitz} \& {Rosolowsky}}{2006}]{blitz.2006} 
  {Blitz} L. \& {Rosolowsky} E. 2006, \apj, 650, 933 


\bibitem[\protect\citeauthoryear{{Bower}, {Benson}, {Malbon}, {Helly}, {Frenk},
    {Baugh}, {Cole} \& {Lacey}}{{Bower} et~al.}{2006}]{bower.etal.2006}
  {Bower} R.~G.,  {Benson} A.~J.,  {Malbon} R.,  {Helly} J.~C.,  {Frenk} C.~S.,
  {Baugh} C.~M.,  {Cole} S. \& {Lacey} C.~G.,  2006, \mnras, 370, 645
  
\bibitem[\protect\citeauthoryear{{Braun}}{{Braun}}{2007}]{braun.2007}
  {Braun} R.,  2007, preprint (arXiv:astro-ph/0703746)
  
\bibitem[\protect\citeauthoryear{{Burke} \& {Graham-Smith}}{{Burke} \&
    {Graham-Smith}}{1996}]{intro.radio.astro}
  {Burke} B.~F. \&  {Graham-Smith} F.,  1996, \emph{An Introduction to Radio
    Astronomy}, Cambridge University Press.
  
\bibitem[\protect\citeauthoryear{{Chengalur}, {Braun} \&
    {Wieringa}}{{Chengalur} et~al.}{2001}]{2001A&A...372..768C}
  {Chengalur} J.~N.,  {Braun} R. \&  {Wieringa} M.,  2001, \aap, 372, 768
  
\bibitem[\protect\citeauthoryear{{Cole}, {Lacey}, {Baugh} \& {Frenk}}{{Cole}
    et~al.}{2000}]{cole.etal.2000}
  {Cole} S.,  {Lacey} C.~G., {Baugh} C.~M. \& {Frenk} C.~S.,  2000, \mnras,
  319, 168
  
\bibitem[\protect\citeauthoryear{{Cole et al.}}{{Cole et al.}}{2001}]{cole.etal.2001}
  {Cole} S. et al., 2001, \mnras, 326, 255 

\bibitem[\protect\citeauthoryear{{Cole et al.}}{{Cole et al.}}{2008}]{cole.etal.2007} 
  Cole, S., Helly, J., Frenk, C.~S., \& Parkinson, H.\ 2008, \mnras, 383, 546 

\bibitem[\protect\citeauthoryear{{Croton}, {Springel}, {White}, {De Lucia},
    {Frenk}, {Gao}, {Jenkins}, {Kauffmann}, {Navarro} \& {Yoshida}}{{Croton}
    et~al.}{2006}]{croton.etal.2006}
  {Croton} D.~J.,  {Springel} V.,  {White} S.~D.~M.,  {De Lucia} G.,  {Frenk}
  C.~S.,  {Gao} L.,  {Jenkins} A.,  {Kauffmann} G.,  {Navarro} J.~F. \&
  {Yoshida} N.,  2006, \mnras, 365, 11
  
\bibitem[\protect\citeauthoryear{{Davis et al.}}{{Davis et al.}}{1985}]{davis.etal.1985} 
  Davis, M., Efstathiou, G., Frenk, C.~S., \& White, S.~D.~M.\ 1985, \apj, 292, 371 

\bibitem[\protect\citeauthoryear{{De Lucia} \& {Blaizot}}{{De Lucia} \&
    {Blaizot}}{2007}]{delucia.blaizot.2007}
  {De Lucia} G. \&  {Blaizot} J.,  2007, \mnras, 375, 2
  
\bibitem[\protect\citeauthoryear{{De Lucia}, {Springel}, {White}, {Croton} \&
    {Kauffmann}}{{De Lucia} et~al.}{2006}]{delucia.etal.2006}
  {De Lucia} G.,  {Springel} V.,  {White} S.~D.~M.,  {Croton} D. \& {Kauffmann}
  G.,  2006, \mnras, 366, 499
  
  
\bibitem[\protect\citeauthoryear{{Drory}, {Salvato}, {Gabasch}, {Bender},
    {Hopp}, {Feulner} \& {Pannella}}{{Drory} et~al.}{2005}]{2005ApJ...619L.131D}
  {Drory} N.,  {Salvato} M.,  {Gabasch} A.,  {Bender} R.,  {Hopp} U.,  {Feulner}
  G. \&    {Pannella} M.,  2005, \apjl, 619, L131
  
\bibitem[\protect\citeauthoryear{{Elmegreen}}{{Elmegreen}}{1993}]{elmegreen.1993} 
  {Elmegreen} B.~G. 1993, \apj, 411, 170 

\bibitem[\protect\citeauthoryear{{Ferri{\`e}re}}{{Ferri{\`e}re}}{2001}]{ferriere.2001} 
  {Ferri{\`e}re} K.~M. 2001, Reviews of Modern Physics, 73, 1031 

\bibitem[\protect\citeauthoryear{{Font}, {Bower}, {McCarthy}, {Benson},
    {Frenk}, {Helly}, {Lacey}, {Baugh} \& {Cole}}{{Font}
    et~al.}{2008}]{font.etal.2008}
  {Font} A.~S.,  {Bower} R.~G.,  {McCarthy} I.~G.,  {Benson} A.~J.,  {Frenk}
  C.~S.,  {Helly} J.~C.,  {Lacey} C.~G.,  {Baugh} C.~M. \&    {Cole} S.,  2008,
  \mnras, 389, 1619
  
\bibitem[\protect\citeauthoryear{{Fontana} et~al.}{{Fontana} et~al.}{2004}]{2004A&A...424...23F}
  {Fontana} A. et~al.,  2004, \aap, 424, 23
  
\bibitem[\protect\citeauthoryear{{Giovanelli et al.}}{{Giovanelli et
      al.}}{2005}]{giovanelli.etal.2005}
  {Giovanelli} R. et al.,  2005, \aj, 130, 2598
  
\bibitem[\protect\citeauthoryear{{Gonzalez}, {Lacey}, {Baugh}, {Frenk} \&
    {Benson}}{{Gonzalez} et~al.}{2009}]{gonzalez.etal.2008}
  {Gonzalez} J.~E.,  {Lacey} C.~G.,  {Baugh} C.~M.,  {Frenk} C.~S. \&  {Benson}
  A.~J., 2009, \mnras, 397, 1254 

  
\bibitem[\protect\citeauthoryear{{Gonzalez-Perez}, {Baugh}, {Lacey} \&
    {Almeida}}{{Gonzalez-Perez} et~al.}{2009}]{gonzalez.etal.2009}
  {Gonzalez-Perez} V.,  {Baugh} C.~M.,  {Lacey} C.~G. \&  {Almeida} C.,  2009,
  MNRAS, 398, 497
  
\bibitem[\protect\citeauthoryear{{Haffner et al}}{{Haffner et al}}{2009}]{haffner.etal.2009} 
  Haffner, L.~M. et al., 2009, Reviews of Modern Physics, 81, 969 

\bibitem[\protect\citeauthoryear{{Harker et al}}{{Harker et al}}{2006}]{harker.etal.2006} 
  Harker, G., Cole, S., Helly, J., Frenk, C., \& Jenkins, A.\ 2006, \mnras, 367, 1039 

\bibitem[\protect\citeauthoryear{{Helly et al.}}{{Helly et al.}}{2003}]{helly.etal.2003} 
  Helly, J.~C., Cole, S., Frenk, C.~S., Baugh, C.~M., Benson, A., \& Lacey, C.\ 2003, \mnras, 338, 903 

\bibitem[\protect\citeauthoryear{{Hopkins}}{{Hopkins}}{2004}]{hopkins.2004}
  Hopkins, A.~M.\ 2004, \apj, 615, 209 

\bibitem[\protect\citeauthoryear{{Jenkins et al.}}{{Jenkins et al.}}{2001}]{jenkins.etal.2001}
  Jenkins, A., Frenk, C.~S., White, S.~D.~M., Colberg, J.~M., Cole, S., Evrard, A.~E., Couchman, H.~M.~P., \& Yoshida, N.\ 2001, \mnras, 321, 372 

\bibitem[\protect\citeauthoryear{{Johnston et al.}}{{Johnston et al.}}{2008}]{askap.science.2008}
  {Johnston} S. et al.,  2008, Experimental Astronomy, 22, 151
  
\bibitem[\protect\citeauthoryear{{Jonas}}{{Jonas}}{2007}]{meerkat.2007}
  {Jonas} J.,  2007, in ``From Planets to Dark Energy: the Modern Radio 
  Universe'', October 1-5 2007, The University of Manchester, UK.
  
\bibitem[\protect\citeauthoryear{{Kennicutt} Jr.}{{Kennicutt}}{1998}]{1998ApJ...498..541K}
  {Kennicutt} Jr. R.~C.,  1998, \apj, 498, 541
  
\bibitem[\protect\citeauthoryear{{Keres}, {Yun} \& {Young}}{{Keres}
    et~al.}{2003}]{keres.etal.2003}
  {Keres} D.,  {Yun} M.~S. \&  {Young} J.~S.,  2003, \apj, 582, 659

\bibitem[\protect\citeauthoryear{{Kilborn} et~al.}{{Kilborn} et~al.}{2009}]{kilborn.etal.2009}
  {Kilborn} V.~A., Forbes, D.~A., Barnes, D.~G., Koribalski, B.~S., Brough, S., 
  \& Kern, K.\ 2009, \mnras, 400, 1962 
  
\bibitem[\protect\citeauthoryear{{Kitzbichler} \& {White}}{{Kitzbichler} \&
    {White}}{2007}]{kitzbichler.white.2007}
  {Kitzbichler} M.~G. \& {White} S.~D.~M.,  2007, \mnras, 376, 2
  
\bibitem[\protect\citeauthoryear{{Krumholz}, {McKee} \& {Tumlinson}}{{Krumholz} et al.}{2009}]{krumholz.etal.2009} 
  {Krumholz} M.~R., {McKee} C.~F. \& {Tumlinson} J. 2009, \apj, 693, 216 

\bibitem[\protect\citeauthoryear{{Lacey} et~al.}{{Lacey} et~al.}{2008}]{lacey.etal.2008} 
  {Lacey, C.~G., Baugh, C.~M., Frenk, C.~S., Silva, L., Granato, G.~L., \& Bressan, A.\ 2008, \mnras, 385, 1155}
  
\bibitem[\protect\citeauthoryear{{Lah} et al.}{{Lah}
    et~al.}{2007}]{lah.etal.2007}
  {Lah} P., et al.,  2007, \mnras, 376, 1357
  
\bibitem[\protect\citeauthoryear{{Lah}, {Pracy}, {Chengalur}, {Briggs},
    {Colless}, {De Propris}, {Ferris}, {Schmidt} \& {Tucker}}{{Lah}
    et~al.}{2009}]{lah.etal.2009}
  {Lah} P.,  {Pracy} M.~B.,  {Chengalur} J.~N.,  {Briggs} F.~H.,  {Colless} M.,
  {De Propris} R.,  {Ferris} S.,  {Schmidt} B.~P. \&  {Tucker} B.~E.,  2009,
  MNRAS, 399, 1447

\bibitem[\protect\citeauthoryear{{Leroy} et al.}{{Leroy} et al.}{2008}]{leroy.etal.2008} 
  Leroy A.~K., Walter F., Brinks E., Bigiel F., de Blok W.~J.~G., 
    Madore~B. \& Thornley, M.~D., 2008, \aj, 136, 2782 
 
\bibitem[\protect\citeauthoryear{{Madau}, {Ferguson}, {Dickinson},
    {Giavalisco}, {Steidel} \& {Fruchter}}{{Madau}
    et~al.}{1996}]{madau.etal.1996}
  {Madau} P.,  {Ferguson} H.~C.,  {Dickinson} M.~E.,  {Giavalisco} M.,
  {Steidel} C.~C. \&  {Fruchter} A.,  1996, \mnras, 283, 1388

\bibitem[\protect\citeauthoryear{{McCarthy}, {Frenk}, {Font}, {Lacey}, {Bower},
    {Mitchell}, {Balogh} \& {Theuns}}{{McCarthy}
    et~al.}{2008}]{mccarthy.etal.2008}
  {McCarthy} I.~G.,  {Frenk} C.~S.,  {Font} A.~S.,  {Lacey} C.~G.,  {Bower}
  R.~G.,  {Mitchell} N.~L.,  {Balogh} M.~L. \& {Theuns} T.,  2008, \mnras,
  383, 593
  
\bibitem[\protect\citeauthoryear{{Meyer et al.}}{{Meyer et al.}}{2004}]{meyer.etal.2004}
  {Meyer} M.~J. et al.,  2004, \mnras, 350, 1195

\bibitem[\protect\citeauthoryear{{Obreschkow} \& {Rawlings}}{{Obreschkow} \&
    {Rawlings}}{2009}]{obreschkow2009a}
  {Obreschkow} D. \& {Rawlings} S.,  2009, \mnras, 394, 1857 

\bibitem[\protect\citeauthoryear{{Obreschkow} et al.}{{Obreschkow} et~al.}{2009a}]{obreschkow2009b}
  {Obreschkow} D.,  {Croton} D.,  {De Lucia} G.,  {Khochfar} S. \& {Rawlings}
  S.,  2009a, \apj, 698, 1467 

\bibitem[\protect\citeauthoryear{{Obreschkow} et al.}{{Obreschkow}
    et~al.}{2009b}]{obreschkow2009c} 
  Obreschkow, D., Heywood, I., Kl\"ockner, H.~-R., \& Rawlings, S., 2009b,
  ApJ, 703, 1890


\bibitem[\protect\citeauthoryear{{Parkinson}, {Cole} \& {Helly}}{{Parkinson}
    et~al.}{2008}]{parkinson.2008}
  {Parkinson} H.,  {Cole} S. \& {Helly} J.,  2008, \mnras, 383, 557
  
\bibitem[\protect\citeauthoryear{{P{\'e}roux}, {McMahon}, {Storrie-Lombardi} \&
    {Irwin}}{{P{\'e}roux} et~al.}{2003}]{peroux.etal.2003}
  {P{\'e}roux} C.,  {McMahon} R.~G.,  {Storrie-Lombardi} L.~J. \& {Irwin} M.~J.,
  2003, \mnras, 346, 1103
 
\bibitem[\protect\citeauthoryear{{Prochaska}, {Herbert-Fort} \&
    {Wolfe}}{{Prochaska} et~al.}{2005}]{prochaska.etal.2005}
  {Prochaska} J.~X.,  {Herbert-Fort} S. \& {Wolfe} A.~M.,  2005, \apj, 635, 123
  
\bibitem[\protect\citeauthoryear{{Quilis}, {Moore} \& {Bower}}{{Quilis}
    et~al.}{2000}]{quilis.etal.2000}
  {Quilis} V.,  {Moore} B. \& {Bower} R.,  2000, Science, 288, 1617
  
\bibitem[\protect\citeauthoryear{{Rao} \& {Briggs}}{{Rao} \&
    {Briggs}}{1993}]{rao.and.briggs.1993}
  {Rao} S. \&  {Briggs} F.,  1993, \apj, 419, 515
  
\bibitem[\protect\citeauthoryear{{Rao}, {Turnshek} \& {Nestor}}{{Rao}
    et~al.}{2006}]{rao.etal.2006}
  {Rao} S.~M.,  {Turnshek} D.~A. \& {Nestor} D.~B.,  2006, \apj, 636, 610

\bibitem[\protect\citeauthoryear{{Reynolds}}{{Reynolds}}{2004}]{reynolds.2004}
  {Reynolds} R.~J., 2004, Advances in Space Research, 34, 1
  
\bibitem[\protect\citeauthoryear{{Spitzer}}{{Spitzer}}{1978}]{spitzer1978}
  {Spitzer} L.,  1978, \emph{Physical Processes in the Interstellar Medium},
  New York Wiley-Interscience.

\bibitem[\protect\citeauthoryear{{Springel et al.}}{{Springel et al.}}{2001}]{springel.etal.2001}
  Springel, V., White, S.~D.~M., Tormen, G., \& Kauffmann, G.\ 2001, \mnras, 328, 726 

\bibitem[\protect\citeauthoryear{{Springel}, {White}, {Jenkins}, {Frenk},
    {Yoshida}, {Gao}, {Navarro}, {Thacker}, {Croton}, {Helly}, {Peacock}, {Cole}, {Thomas}, {Couchman}, {Evrard}, {Colberg} \& {Pearce}}{{Springel}
    et~al.}{2005}]{springel.etal.2005}
  {Springel} V.,  {White} S.~D.~M.,  {Jenkins} A.,  {Frenk} C.~S.,  
  {Yoshida} N.,
  {Gao} L.,  {Navarro} J.,  {Thacker} R.,  {Croton} D.,  {Helly} J.,
  {Peacock} J.~A.,  {Cole} S.,  {Thomas} P.,  {Couchman} H.,  {Evrard} A.,
  {Colberg} J. \& {Pearce} F.,  2005, \nat, 435, 629

\bibitem[\protect\citeauthoryear{{Springob} et~al.}{{Springob} et~al.}{2008}]{springob.etal.2005} 
  Springob, C.~M., Haynes, M.~P., \& Giovanelli, R.\ 2005, \apj, 621, 215 

\bibitem[\protect\citeauthoryear{{Taylor}}{{Taylor}}{2008}]{taylor.2008} 
  Taylor, A.~R.\ 2008, IAU Symposium, 248, 164 

\bibitem[\protect\citeauthoryear{{Verheijen}, {Oosterloo}, {van Cappellen},
    {Bakker}, {Ivashina} \& {van der Hulst}}{{Verheijen}
    et~al.}{2008}]{apertif.2008}
  {Verheijen} M.~A.~W.,  {Oosterloo} T.~A.,  {van Cappellen} W.~A.,  
  {Bakker} L.,
  {Ivashina} M.~V. \& {van der Hulst} J.~M.,  2008, in {Minchin} R.,
  {Momjian} E.,  eds, ``The Evolution of Galaxies Through the Neutral Hydrogen
  Window'' Vol.~1035 of American Institute of Physics Conference Series.,
  
\bibitem[\protect\citeauthoryear{{Weinmann}, {van den Bosch}, {Yang} \&
    {Mo}}{{Weinmann} et~al.}{2006a}]{weinmann.etal.2006a}
  {Weinmann} S.~M.,  {van den Bosch} F.~C.,  {Yang} X. \& {Mo} H.~J.,  2006a,
  \mnras, 366, 2
  
\bibitem[\protect\citeauthoryear{{Weinmann}, {van den Bosch}, {Yang}, {Mo},
    {Croton} \& {Moore}}{{Weinmann} et~al.}{2006b}]{weinmann.etal.2006b}
  {Weinmann} S.~M.,  {van den Bosch} F.~C.,  {Yang} X.,  {Mo} H.~J.,  {Croton}
  D.~J. \& {Moore} B.,  2006b, \mnras, 372, 1161

\bibitem[\protect\citeauthoryear{Wong \& Blitz}{2002}]{wong2002} 
  Wong T., Blitz L., 2002, ApJ, 569, 157 

\bibitem[\protect\citeauthoryear{{Young \& Scoville}}{{Young \& Scoville}}{1991}]{young.scoville.91}
  {Young} J.~S. \& {Scoville} N.~Z., 1991, \araa, 29, 581
  
\bibitem[\protect\citeauthoryear{{Zwaan}}{{Zwaan}}{2000}]{zwaan.2000}
  {Zwaan} M.~A.,  2000, PhD thesis, PhD Thesis, Groningen
  
\bibitem[\protect\citeauthoryear{{Zwaan}, {Briggs}, {Sprayberry} \&
    {Sorar}}{{Zwaan} et~al.}{1997}]{zwaan.etal.1997}
  {Zwaan} M.~A.,  {Briggs} F.~H.,  {Sprayberry} D. \& {Sorar} E.,  1997, \apj,
  490, 173
  
\bibitem[\protect\citeauthoryear{{Zwaan et al.}}{{Zwaan et al.}}{2003}]{zwaan.etal.2003}
  {Zwaan} M.~A. et al.,  2003, \aj, 125, 2842

\bibitem[\protect\citeauthoryear{{Zwaan}, {Meyer}, {Staveley-Smith} \&
    {Webster}}{{Zwaan} et~al.}{2005}]{zwaan.etal.2005}
  {Zwaan} M.~A.,  {Meyer} M.~J.,  {Staveley-Smith} L. \& {Webster} R.~L.,  2005,
  \mnras, 359, L30
  
\end{thebibliography}

\label{lastpage}

\end{document}